\documentclass[11pt,a4paper]{article}

\usepackage[ngerman,english]{babel}
\usepackage{mathtools}
\usepackage{jheppub}
\usepackage{dsfont}
\usepackage{amsfonts}
\usepackage{amsmath}
\usepackage{amssymb}
\usepackage{tabularx}
\usepackage{makecell}
\usepackage{multirow}
\usepackage{siunitx}
\usepackage{placeins}
\usepackage{hyperref}
\hypersetup{linktoc=all,hidelinks}
	\addto\extrasenglish{

	  }
\usepackage{caption}

\newcommand*{\db}[2][]{\!\mathrm{d}^{#1}#2\!\mathop{}}
\newcommand{\inv}[1]{\frac{1}{#1}}
\newcommand*{\SU}[1]{SU($#1$)}

\newcommand*{\ldef}{\mathrel{\vcenter{\baselineskip0.5ex \lineskiplimit0pt\hbox{\scriptsize.}\hbox{\scriptsize.}}}
	\hspace{-3.5pt}\scalebox{0.8}[1]{$=$}~}

\newcommand{\imagi}{\mathrm{i}}

\DeclareMathOperator{\tr}{Tr}
\DeclareMathOperator{\re}{Re}
\DeclareMathOperator{\im}{Im}
\DeclareMathOperator{\atanh}{atanh}

\DeclareMathOperator{\hc}{h.c.}

\def\bx{{\mathbf{x}}}

\def\bp{{\mathbf{p}}}

\title{From the colour glass condensate to filamentation: Systematics of
classical Yang-Mills theory}
\author[a,b]{Owe Philipsen,}
\author[a,b]{Bj\"orn Wagenbach,}
\author[c]{Savvas Zafeiropoulos}

\affiliation[a]{Institut f\"ur Theoretische Physik, Johann Wolfgang
Goethe-Universit\"at,\\Max-von-Laue-Str.~1, 60438 Frankfurt am Main, Germany}
\affiliation[b]{John von Neumann Institute for Computing (NIC) GSI,\\Planckstr.~1, 
64291 Darmstadt, Germany}
\affiliation[c]{Institut f\"ur Theoretische Physik, Universit\"at Heidelberg,\\Philosophenweg 12, 69120 Heidelberg,
Germany}

\emailAdd{philipsen@th.physik.uni-frankfurt.de}
\emailAdd{wagenbach@th.physik.uni-frankfurt.de}
\emailAdd{zafeiropoulos@thphys.uni-heidelberg.de}

\abstract{
The non-equilibrium early time evolution of an ultra-relativistic
heavy ion collision is often described by classical 
lattice Yang-Mills theory, starting from the colour glass condensate (CGC) effective theory
with an anisotropic energy momentum tensor as initial condition.
In this work we investigate the systematics associated with such studies
and their dependence on various model parameters (IR, UV cutoffs and the
amplitude of quantum fluctuations) which are not yet fixed by experiment.
We perform calculations for \SU{2} and \SU{3}, both in a static box and in an 
expanding geometry. Generally, the dependence on model parameters is found to 
be much larger than that on technical parameters like the number of colours, boundary conditions 
or the lattice spacing.
In a static box, all setups lead to isotropisation
through chromo-Weibel instabilities, which is illustrated by the accompanying
filamentation of the energy density. However, the associated time scale depends strongly
on the model parameters and in all cases is longer than the phenomenologically expected one.
In the expanding system, no isotropisation is observed for any parameter choice.
We show how investigations at fixed initial energy density can be used to better
constrain some of the model parameters.
}

\begin{document}

\maketitle

\section{Introduction}\label{sec:intro}

The medium created by ultra-relativistic heavy-ion collisions is characterised by strong 
collective behaviour. It is generally accepted that a quark-gluon plasma (QGP) is
formed and the effective theory describing the 
multiparticle correlations of this nearly-perfect fluid is relativistic viscous hydrodynamics.
The application of hydrodynamic models requires the thermalisation time scale from the 
initial non-equilibrium stage of the collision to the QGP to be very 
fast ~\cite{Heinz:2001xi,Romatschke:2007mq} compared to the lifetime of the QGP.

From a theoretical point of view, a heavy-ion collision has different stages. 
As an initial condition, one assumes the colour glass condensate (CGC), 
i.e.~an effective field theory description of boosted, saturated gluons \cite{Iancu2001}. 
The resulting strong gauge field dynamics constitutes the first stage of the 
evolution. The following 
second stage is then governed by hydrodynamic equations until the medium 
becomes too dilute for this long wavelength description. The precise duration of stage I 
is not yet known for realistic values of the coupling. 
Hydrodynamic models of stage II constrain it to be around or less 
than $\SI{1}{\femto\meter}$ \cite{Kolb2000}.

The evolution of soft gauge fields during stage I, including
dynamical instabilities such as the chromo-Weibel 
instability \cite{Heinz:1985vf,Mrowczynski:1988dz,Pokrovsky:1988bm,Mrowczynski:1993qm,Blaizot:2001nr,Romatschke:2003ms,Arnold:2003rq,Carrington:2014bla}, is a subject of intense 
research. Field dynamics in an expanding background has been extensively 
studied using numerical simulations of classical Yang-Mills theory \cite{Romatschke:2005pm,Romatschke2006b,Fukushima2011,Berges:2012iw,Berges:2013eia,Fukushima:2013dma,Gelis2013a}, 
perturbative approaches in the high energy limit \cite{Kurkela:2011ti,Kurkela:2011ub}, 
and kinetic \SU{2} Vlasov-Yang-Mills equations 
\cite{Romatschke:2006wg,Rebhan:2008uj,Attems:2012js}.

In this work, we focus on the early time dynamics of the gauge fields out of equilibrium, 
where we pursue a purely classical treatment of Yang-Mills theory. 
This approach is justified for the infrared modes of gauge fields with a high occupation 
number.

Our goal is to initiate a systematic study of the dependence on a 
variety of parameters entering through the CGC initial condition as well as the
systematics of the classical evolution itself.
In particular, we compare a treatment of the realistic \SU{3} gauge group 
with the more economical \SU{2}, monitor a gauge-invariant definition of the
occupation number of field modes to address the validity of the classical approximation,
and compare the evolution in a static box with the one in an expanding medium.
We also attempt to quantify the dependence of our results on various model parameters 
introduced in the literature, like 
the amplitude of initial boost non-invariant fluctuations, an IR cutoff to emulate
colour neutrality on the scale of nucleons as well as a UV cutoff on 
the initial momentum distribution. 

In the following section we summarise the theoretical framework of our approach 
and give the CGC initial conditions this work is based on. In \autoref{sec:results}, 
we present the numerical results of our simulations, where we extensively 
elaborate on the underlying parameter space of the CGC. We will see that the system is 
highly sensitive to the model parameters and suggest a method to reduce the number of 
free parameters by keeping the system's physical energy density fixed. 
We also present depictions of the filamentation of the energy density 
in position space, which results from initial quantum fluctuations and indicates 
the occurrence of chromo-Weibel instabilities. 
\autoref{sec:conclusion} contains our conclusions and an outlook. 
Some very early stages of this work appeared as a conference proceeding \citep{Attems2016}.

\section{Classical Yang-Mills theory on the lattice}\label{sec:theory}

\subsection{Hamiltonian formulation}\label{sec:hamilton_formulation}

Our starting point is the Yang-Mills action in general coordinates,
\begin{align}\label{eq:S_gc}
	S = \int \db[4]{x} \mathcal{L} &= -\frac{1}{2} \int \db[4]{x} \sqrt{-\det [(g_{\mu\nu})]} ~ \tr \big[F_{\mu\nu} g^{\mu\alpha} g^{\nu\beta} F_{\alpha\beta} \big]\,.
\end{align}
For a treatment on an 
anisotropic, hypercubic lattice in Minkowski spacetime we employ Wilson's 
formulation\footnote{Unless stated differently, we use the following index convention throughout this paper in order to minimise redundancy: $\mu=0,1,2,3=(t,x,y,z|\tau,x,y,\eta)$, $i=1,2,3=(x,y,z|x,y,\eta)$, $k=1,2=x,y$ and $\zeta=3=(z|\eta)$.} 
\begin{align}
	S=\frac{\beta}{N_c}{\rm Re}{\rm Tr}\left\{\xi\sum_{x,i}(1-U_{ti}(x))
	-\frac{1}{\xi}\sum_{x,i<j}(1-U_{ij}(x))
	\right\} \,.
\end{align}
The anisotropy parameter $\xi=a_\sigma /a_t$ is the ratio of spatial and temporal 
lattice spacings which does not renormalise in the classical limit, and 
$\beta=2N_c/g^2$ is the lattice gauge coupling (we choose $N_c=2$ and $N_c=3$ colours).

In the expanding geometry, where we use comoving coordinates $\tau=\sqrt{t^2-z^2}$ and $\eta=\atanh(z/t)$, the lattice action reads
\begin{align}
	S = \frac{2}{g^2} \sum_x \re\tr \bigg[ \frac{a_\eta\tau}{a_\tau} \sum_k (1-U_{\tau k}) &+ \frac{a_\perp^2}{a_\tau a_\eta \tau} (1-U_{\tau\eta}) \nonumber\\
	&- \frac{a_\eta a_\tau \tau}{a_\perp^2} (1-U_{xy}) - \frac{a_\tau}{a_\eta \tau} \sum_k (1-U_{k\eta}) \bigg] \,.
\end{align}
We introduced the transverse lattice spacing $a_\perp$ and the dimensionless rapidity discretisation $a_\eta$. Inserting the link variables
\begin{align}
	U_\mu(x) = e^{\imagi ga_\mu A_\mu(x)}
\end{align}
into the plaquettes $U_{\mu\nu}(x) \equiv U_\mu(x) U_\nu(x+\hat{\mu}) U_{-\mu}(x+\hat{\mu}+\hat{\nu}) U_{-\nu}(x+\hat{\nu})$, $U_{-\mu}(x) \equiv U_\mu^\dagger(x-\hat{\mu})$, and expanding around small values of the lattice spacing one recovers the classical Yang-Mills action in the continuum limit, $a_\mu\!\rightarrow\!0$. In order to choose canonical field variables and construct a Hamiltonian, we set
\begin{align}
	A_{t/\tau}(x) = 0 \quad \Leftrightarrow \quad U_{t/\tau}(x) = 1\,,
\end{align}
i.e., we are using temporal gauge. 
The field variables are then the spatial (and rapidity) links
\begin{align}
	U_i(x) &= e^{\imagi ga_i A_i(x)}
\end{align}
and the rescaled dimensionless chromo-electric fields,
\begin{subequations}
	\begin{align}
		\text{static box:}\quad E_i(x) &= g a_\sigma^2 \partial_tA_i(x)\,,\\
		\text{expanding system:}\quad E_k(x) &= g a_\perp \tau \partial_\tau A_k(x) \,,\quad E_\eta(x) = ga_\perp^2\frac{1}{\tau} \partial_\tau A_\eta(x)\,.
	\end{align}
\end{subequations}
For the situation in a static box this results in the standard Hamiltonian
\begin{align}
	H[U_i,E_i]=\frac{1}{g^2 a_\sigma}\sum_{\bx}\re\tr\left\{2\sum_{i<j}\big[1-U_{ij}\big]
	+\sum_i E_i^2\right\}\,,
\end{align}
with corresponding classical field equations
\begin{subequations}\label{eq:eom_sb}
\begin{align}
	U_i(x+\hat{t})&=\exp \big[\imagi\xi^{-1}E_i(x)\big]U_i(x)\,,\\
	E_i^a(x+\hat{t})&=E_i^a(x)+2\xi^{-1}\sum_{j\neq i}\im\tr \Big\{T^a\big[U_{ji}(x)+U_{-ji}(x)\big]\Big\}\,,
\end{align}
\end{subequations}
and Gauss constraint
\begin{align}\label{eq:gauss_law_sb}
	G^a(x)=\sum_i \im\tr \Big\{ T^a \big[ U_{it}(x) + U_{-it}(x) \big] \Big\} \bigg|_{U_t=1}=0\;.
\end{align}
For the expanding case we have, in comoving coordinates,
\begin{align}
	H[U_i,E_i]=\frac{a_\eta}{g^2 a_\perp}\sum_{\bx}\re\tr\left\{2\tau\big[1-U_{xy}\big]+\frac{2}{a_\eta^2\tau}\sum_{k}\big[1-U_{k\eta}\big]+\frac{E_k^2}{\tau}+\tau E_\eta^2\right\}\,,
\end{align}
with field equations
\begin{subequations}\label{eq:eom_es}
	\begin{align}
		U_k\,(x+\hat{\tau}) &= \exp \!\bigg[ \imagi\frac{a_\tau}{\tau} E_k(x) \bigg] U_k(x)\,,\\[0.25cm]
		U_\eta(x+\hat{\tau}) &= \exp \!\big[ \imagi a_\eta a_\tau \tau E_\eta(x) \big] U_\eta(x)\,,\\
		E_k^a(x+\hat{\tau}) &= E_k^a(x) + 2 \im\tr \!\bigg\{\! T^a \Big( a_\tau \tau \!\sum_{j \neq k}\! \big[ U_{jk}(x) + U_{-jk}(x) \big] + \frac{a_\tau}{a_\eta^2 \tau} \big[U_{\eta k}(x) + U_{-\eta k}(x) \big] \!\Big) \!\bigg\},\\
		E_\eta^a(x+\hat{\tau}) &= E_\eta^a(x) + \frac{2a_\tau}{a_\eta \tau} \sum_k \im\tr \!\bigg\{ T^a \Big[ U_{k\eta}(x) + U_{-k\eta}(x) \Big] \bigg\}\,,
	\end{align}
\end{subequations}
and Gauss constraint
\begin{align}\label{eq:gauss_law_es}
	G^a(x)=-\im\tr \bigg\{ T^a \bigg( \frac{\tau}{a_\perp} \sum_k \Big[ U_{k\tau}(x) &+ U_{-k\tau}(x) \Big] \nonumber\\
	+ \frac{a_\perp}{a_\eta^2 \tau} \Big[ U_{\eta\tau}(x) &+ U_{-\eta\tau}(x) \Big]\bigg)\bigg\} \bigg|_{U_\tau=1}=0 \,.
\end{align}
We then consider the time evolution of the classical statistical system whose equilibrium 
states are determined by the classical partition function
\begin{align}
	Z=\int DU_iDE_i\;\delta(G)\,e^{-\frac{H}{T}}\,.
\end{align}
For simulations in equilibrium, initial configurations are 
generated with a thermal distribution governed by this partition function, 
and then evolved in $t$ by solving \eqref{eq:eom_sb} or \eqref{eq:eom_es}, respectively. 
For a system out of equilibrium there is no partition function. 
Rather, initial fields satisfying the Gauss constraint have to be specified by 
some initial conditions and are then evolved using the field equations.

\subsection{Non-equilibrium initial conditions (CGC)}\label{sec:initial_conditions}

Heavy-ion collisions at high energy density can be described in terms of deep inelastic scattering
of partons. The corresponding parton distribution functions are dominated by gluonic contributions, which
motivates the description in terms of a colour glass effective theory \cite{Iancu2001,McLerran2002}.
The gluonic contribution to the parton distribution is limited by a saturation momentum $Q_s$, which is
proportional to the collision energy.
When the saturation scale $Q_s$ becomes large there is a time frame where
soft and hard modes get separated \cite{Iancu:2003xm}.
The colliding nuclei constitute hard colour
sources, which can be seen as static. Due to time dilatation,
they are described as thin sheets of colour charge. 

Choosing $z$ as the direction of the collision, this is usually described in light cone 
coordinates,
\begin{align}
	x^{\pm}=\frac{t\pm z}{\sqrt{2}},\quad  \bx_\perp=(x,y)\,.
\end{align}
The colour charges are distributed randomly from collision to collision.  
In the McLerran-Venugopalan (MV) model \cite{McLerran:1993ka} the distribution is taken to be Gaussian, with charge densities ($a,b \in \{1,\dots,N_g\ldef N_c^2-1\}$),
\begin{align}\label{eq:init}
	\big\langle \rho_v^{k,a}(x_\perp) \rho_w^{l,b}(y_\perp) \big\rangle
	&= a_\perp^4 \frac{g^4 \mu^2}{N_l} \delta_{vw} \delta^{kl} \delta^{ab} \delta(x_\perp - y_\perp)\,.
\end{align}
Here $\mu^2\sim A^{1/3}$\,\si{\per\femto\meter\squared} is the colour charge squared 
per unit area in one colliding nucleus
with atomic number $A$. It is non-trivially related to the saturation scale \cite{Lappi2008}, with $Q_s \approx Q \ldef g^2\mu$. For $Pb-Pb$ or $Au-Au$ collisions, this is larger than the fundamental QCD scale $\Lambda_{\rm QCD}$.
We choose a value in the range of expectations for ultra-relativistic heavy-ion
collision at the Large Hadron Collider ($Q_s \approx \SI{2}{\giga\eV}$ \cite{Fujii2009}) 
and fix $Q =  \SI{2}{\giga\eV}$ for our simulations throughout this paper. 

Originally the MV model was formulated for a fixed time slice. Later it was 
realised that, in order to maintain gauge-covariance in the longitudinal direction, 
this initial time slice has to be viewed as a short-time limit of a construction using $N_l$ time slices, 
containing Wilson lines in the longitudinal direction
\cite{Fukushima2008,Lappi2008}. In the literature the designation ''$N_y$`` is also frequently used for the number of longitudinal sheets, but in order to distinguish it from the lattice extent in $y$-direction we use $N_l$ instead.

The colour charge densities produce the non-Abelian current
\begin{align}\label{eq:IC_current}
	J^{\mu,a}(x)=\delta^{\mu +} \rho_{1}^a(x_\perp,x^-)+
	\delta^{\mu -} \rho_{2}^a(x_\perp,x^+)
\end{align}
and the corresponding
classical gluon fields are then obtained by solving the Yang-Mills equations in the presence of
those sources,
\begin{align}
	D_\mu F^{\mu\nu}=J^\nu\,.
\end{align}
For the lattice implementation of this initial condition, we follow \cite{Lappi2008} and solve
\begin{subequations}\label{eq:CGC_IC}
	\begin{align}
		\big[\Delta_L+m^2\big] \Lambda^{k,a}_v(x_\perp) &= - \bar{\rho}^{k,a}_v(x_\perp) \label{eq:Poisson} \\
		V^k(x_\perp) &= \prod_{v=1}^{N_l} \exp\big[\imagi\Lambda_v^k(x_\perp)\big] \label{eq:Vm} \\
		U^k_i(x_\perp) &= V^k(x_\perp){V^k}^\dagger(x_\perp+\hat{\imath}) \label{eq:Um}
	\end{align}
\end{subequations}
with the lattice Laplacian in the transverse plane,
\begin{align}
	\Delta_L\Lambda(x_\perp)=\sum_{i=x,y}\left(\Lambda(x_\perp+\hat{\imath})-2\Lambda(x_\perp)+
	\Lambda(x_\perp-\hat{\imath})\right) \;.
\end{align}
The two nuclei are labelled by $k=1,2$, the index $v=1,\dots,N_l$ indicates 
the transverse slice under consideration
and $m$ is an IR regulator. For $m=0$, a finite lattice volume acts as an effective
IR cutoff. However, a finite $m\sim \Lambda_\mathrm{QCD}$ is expected to exist, since
correlators of colour sources are screened over distances of $\Lambda_\mathrm{QCD}^{-1}$,
as was initially proposed in \cite{Lappi2008}. 
Of course, a determination of this screening length requires the full quantum theory and
thus is beyond a classical treatment. We shall investigate the dependence of our results
by varying $m$ between zero and some value of the expected order of magnitude. 
Physically, the parameter $m$ indicates the inverse length scale over which objects 
are colour neutral in our 
description, and hence 
$m=0.1\,Q\approx\SI{200}{\mega\eV}\approx\SI{1}{\per\femto\meter}\approx\frac{1}{R_p}$, 
with $R_p$ being the proton radius, is a sensible choice.

Although we already have a UV cutoff $\sim\!1/a_\perp$ from the lattice discretisation, 
often an additional UV cutoff $\Lambda$ is used in the literature \cite{Fukushima:2013dma}, 
while solving Poisson's equation \eqref{eq:Poisson}. It can be interpreted as an additional 
model parameter, which restricts the colour sources in Fourier space to modes further in 
the IR. Again, we shall investigate how results depend on the presence and size
of this model parameter. 

To get the transverse components of the collective initial lattice gauge fields $U_k=\exp(\imagi\alpha_a T^a)$, 
$\alpha_a \in \mathds{R}$, we have to solve $N_g$ equations at each point on the 
transverse plane,
\begin{align}\label{eq:initUi}
	\tr \Big\{ T^a \Big[ (U^{(1)}_k + U^{(2)}_k) (1 + U^\dagger_k) - \hc \Big] \Big\} = 0\;.
\end{align}
For the case of $N_c=3$ we do this numerically using multidimensional root finding methods 
of the GSL library \cite{GSL:2009}.
For the case of $N_c=2$, one can find a closed-form expression and circumvent this 
procedure, i.e.~\eqref{eq:initUi} reduces to
\begin{align}
	U_k = \Big(U_k^{(1)} + U_k^{(2)}\Big) \Big({U_k^{(1)}}^\dagger + {U_k^{(2)}}^\dagger\Big)^{-1}\;.
\end{align}
The remaining field components are $U_{\zeta}(x)=1$, $E_k^a(x)=0$ and
\begin{align}
	E_{\zeta}^a(x_\perp) = -\frac{\imagi}{2}\sum_{k=1,2} \tr \bigg\{ T^a \bigg( &\Big[U_k(x) - 1\Big]\Big[U_k^{(2)\dagger}(x) - U_k^{(1)\dagger}(x)\Big]\\
	& + \Big[U_k^\dagger(x-\hat{k}) - 1\Big]\Big[U_k^{(2)}(x-\hat{k}) - U_k^{(1)}(x-\hat{k})\Big] \bigg) - \hc \bigg\}\,,\nonumber
\end{align}
with the index convention introduced in \autoref{sec:hamilton_formulation}.

To make the initial conditions more realistic, fluctuations can be added on top of this 
background \cite{Fukushima:2006ax,Romatschke2006b}, which are
supposed to represent quantum corrections to the purely 
classical fields. They are low momentum modes constructed to satisfy 
the Gauss constraints \eqref{eq:gauss_law_sb} and \eqref{eq:gauss_law_es}, respectively,
\begin{subequations}\label{eq:fluctuations}
	\begin{align}
		\delta E_k^a(x)&=\frac{1}{a_{\zeta}}\big[F_k(x)-F_k(x-\hat{\zeta})\big]\,,\\
		\delta E_\zeta(x)&=-\sum_{k}\left[F_k(x)-U^\dag_k(x-\hat{k})
		F_k(x_\perp-\hat{k})U_k(x_\perp-\hat{k})\right],\\
		F_k(x)&=\Delta\,\cos\left(\frac{2\pi \zeta}{L_\zeta}\right) \chi_k(x_\perp),
	\end{align}
\end{subequations}
where $\chi_k(x_\perp)$ are standard Gaussian distributed random variables on the 
transverse plane. The amplitude of the fluctuations is parametrised by $\Delta$. 
Since there is no theoretical prediction for its value, it is yet another model parameter
which we shall vary in order to study its effect on the physical results.

\subsection{Setting the lattice scale and size}\label{sec:scale_setting}

In a non-equilibrium problem, a scale is introduced by the physical quantity specifying the initial condition. In our case this is the magnitude of the initial colour charge distribution defined in \eqref{eq:init} and we follow again \cite{Fukushima:2013dma} in setting the dimensionless combination $Q L=120$, where $L$ corresponds to the transversal box length in physical units. It is chosen to correspond to the diameter of an $Au$ atom with $A=197$, $R_A=1.2\,A^{1/3}$\,\si{\femto\meter} $\approx \SI{7}{\femto\meter}$. In the LHC literature it is conventional to define the transverse section of the box by $\pi R_A^2=L^2$, which then sets the transverse lattice spacing through $L=N_\perp a_\perp$. Together with $Q=\SI{2}{\giga\electronvolt}$ we thus have
\begin{align}\label{eq:a_perp}
	a_\perp=\frac{L}{N_\perp}=\frac{120}{Q N_\perp}\approx \frac{12}{N_\perp} \si{\femto\meter}\,.
\end{align}
As long as we do not add any term describing quantum fluctuations, the system reduces to a 2D problem and thus the results are independent of $a_\zeta$. For non-vanishing fluctuations in the static box we work with an isotropic spatial lattice, i.e.~$a_z=a_\perp$, whereas our 3D simulations in comoving coordinates are performed at $a_\eta N_\eta = 2.0$ as proposed, e.g., in \cite{Fukushima2012}.

\subsection{Observables}

Energy density and pressure
are convenient observables to investigate the early isotropisation process 
of the plasma. The system's energy density is the 0th diagonal element of the 
energy-momentum tensor, $T^{00}$, and can be separated into its evolving 
chromo-magnetic and chromo-electric components, $\epsilon_B$
and $\epsilon_E$, respectively, 
and further into transverse and longitudinal components,
\begin{align}\label{eq:edens}
	\epsilon=\epsilon_T+\epsilon_L = &
	\epsilon_{B_T}+\epsilon_{E_T}+\epsilon_{B_L}+\epsilon_{E_L}\,.
\end{align}
On the lattice, the chromo-electric and chromo-magnetic contributions to the Hamiltonian density in Cartesian coordinates, $\mathcal{H} \equiv T^{tt}$, are
\begin{subequations}
	\begin{align}
		a_\sigma^4 \mathcal{H}_i^E(t,x) &=
		\frac{\beta}{2 N_c}
		\tr \big[E_i(x)E_i(x)\big] , \\
		a_\sigma^4 \mathcal{H}_i^B(t,x) &=
		\beta
		\sum_{i<j}
		\left[ 1 - \inv{N_c} \re \tr U_{ij}(x) \right]\,.
	\end{align}
\end{subequations}
The contributions to the lattice Hamiltonian density in comoving coordinates, $\mathcal{H} \equiv \tau T^{\tau\tau}$, read
\begin{subequations}
	\begin{align}
		\frac{a_\perp^4}{\tau}\mathcal{H}^E_k(x) &= \frac{\beta a_\perp^2}{2N_c\tau^2} \tr \big[E_k(x) E_k(x)\big]\,, \\
		\frac{a_\perp^4}{\tau}\mathcal{H}^E_\eta(x) &= \frac{\beta}{2N_c} \tr \big[E_\eta(x) E_\eta(x)\big]\,, \\
		\frac{a_\perp^4}{\tau}\mathcal{H}^B_k(x) &= \frac{\beta a_\perp^2}{2N_c a_\eta^2\tau^2} \re\tr\big[ 1-U_{k\eta} \big]\,, \\
		\frac{a_\perp^4}{\tau}\mathcal{H}^B_\eta(x) &= \frac{\beta}{2N_c} \re\tr\big[1-U_{12}\big]\,.
	\end{align}
\end{subequations}
Summing the transverse and longitudinal components over the lattice then gives the averaged energy density contributions,
\begin{subequations}\label{eq:edens_comp}
	\begin{align}
		\epsilon_{E_T} (t) &= \inv{V} \sum_\bx \sum_k\mathcal{H}_k^E(x) \,, \\
		\epsilon_{E_L} (t) &= \inv{V} \sum_\bx \mathcal{H}_\zeta^E(x)\,,\\
		\epsilon_{B_T} (t) &= \inv{V} \sum_\bx \sum_k\mathcal{H}_k^B(x) \,, \\
		\epsilon_{B_L} (t) &= \inv{V} \sum_\bx \mathcal{H}_\zeta^B(x) \,,
	\end{align}
\end{subequations}
with the lattice volume $V = N^2_\perp N_\zeta$.

A suitable measure for isotropisation is given by the ratio of longitudinal and transverse pressure.
These are given by the spatial diagonal elements of the energy momentum tensor, 
\begin{subequations}
	\begin{align}\label{eq:pressures}
		P_T(x)=\frac{1}{2}\big[T^{xx}&(x)+T^{yy}(x)\big]=\epsilon_L \,,\\
		P_L(x)=T^{zz}(x)=\epsilon_T - \epsilon_L \quad &\Big| \quad P_L(x)=\tau^2T^{\eta\eta}(x)=\epsilon_T - \epsilon_L\,.
	\end{align}
\end{subequations}
Note that at early times the field component of the longitudinal pressure
is negative. This is due to the leading order of the CGC initial condition which sets $P_L$ to
exactly the negative value of $P_T$~\cite{Lappi:2006fp},
\begin{align}
	T^{\mu\nu}_{\rm CGC, LO} = {\rm diag} (\epsilon, \epsilon,
	\epsilon, -\epsilon)
\end{align}
and reflects the force of the colliding nuclei. In complete equilibrium both pressures are equal.

\subsection{Validity of the classical approximation}

One requirement for a quantum field to behave effectively 
classically is a high occupation number $N$ of its field modes. 
In addition, for a classical description to be a good approximation, the IR sector should
dominate the total energy of the system, since the classical theory breaks down in the UV.

Thus, in order to study the population of the different momentum modes, it is customary 
in the literature to compute the Fourier components of the chromo-electric field and 
their contribution 
to the energy. However, chromo-electric fields and their Fourier modes are gauge-dependent. 
Besides the ambiguities this causes in the interpretation of the momentum distribution, 
it also introduces a significant computational overhead for the process of gauge fixing. 
For this reason we consider the spectral decomposition for a manifestly gauge-invariant 
quantity, the Fourier transform of the total energy density,
\begin{align}\label{eq:fourier_H}
	\mathcal{H}(t,\bp) = \inv{V} \sum_\bx e^{-i\bp\bx}\sum_i\big[\mathcal{H}_i^E(x)+\mathcal{H}_i^B(x)\big]\,,
\end{align}
whose average over equal absolute values of momenta\footnote{This means we average over 
all vectors with the same length, i.e., all combinations of $\bp=(p_x,p_y,p_z)$ that 
result in the same absolute value $p\equiv|\bp|=\sqrt{p_x^2+p_y^2+p_z^2}$. 
This is indicated by the notation $\langle\,\cdot\,\rangle_p$.} 
normalised on that momentum, provides a measure for the population of momentum modes. 
That is, we define the occupation number density
\begin{align}\label{eq:modes}
	n(p) \ldef \frac{N(p)}{\mathcal{V}} \ldef \frac{1}{p} \,\big\langle |\epsilon(p)| \big\rangle_p \,,
\end{align}
with the physical volume $\mathcal{V}$ and $\epsilon(p) \equiv \mathcal{H}(p)$ in the static case and $\epsilon(p) \equiv \mathcal{H}(p)/\tau$ in the expanding one, respectively.
This definition is motivated by considering a gas of free gluons 
\cite{Krasnitz2001,Lappi2003,Bodeker2007}, where the system's energy 
$\mathcal{E}$ (in $d$ dimensions) is related to the gluon mode density $n$ via
\begin{align}
	\mathcal{E} \equiv \int \text{d}^dp \, \epsilon(\bp) = \int \text{d}^dp \, \omega(\bp) n(\bp)\,,
\end{align}
and the eigenfrequency $\omega$ corresponds to the free massless dispersion 
relation $\omega(p)\approx p$ for $p \gtrsim 0.1\,Q$ \cite{Krasnitz2001}.

With a gauge-invariant definition, the occupation number per mode can be determined
fairly uniquely. A much harder question is
up to which energy level the modes of a classical theory provide a good approximation: 
because of the Rayleigh-Jeans divergence,
the UV sector of the classical theory theory in equilibrium 
increasingly deviates from that of the full quantum theory, irrespective of 
occupation numbers. In thermal equilibrium, a UV cutoff is usually fixed by matching a
thermodynamical observable between the full and an effective theory. In 
a non-equilibrium situation, however, it is difficult to identify a  
scale up to which the classical
theory is valid. A common self-consistent procedure then is to demand that the total energy of the system under 
study is ``dominated by infrared modes''.    

\subsection{Ordering of scales and parameters}

We wish to study the dependence of the classical Yang-Mills system 
on the lattice spacing and volume, as well
as of the various parameters introduced through the CGC initial conditions. 
For the classical description of the CGC model to be self-consistent, the
parameters representing various scales of the problem have to satisfy
\begin{align}\label{eq:CGC_params}
	\frac{1}{L_\perp} ~\lesssim~ m ~\ll~ Q ~\ll~ \Lambda ~\lesssim~ \frac{1}{a_\perp}\;.
\end{align}
The original MV model without additional IR and UV cutoffs corresponds to the special case
$m=L_\perp^{-1}$ and $\Lambda=a_\perp^{-1}$.  
The dimensionless version of these relations to be satisfied by our lattice simulation 
is obtained by dividing everything by $Q$.

\section{Numerical results}\label{sec:results}

Our numerical implementation is based on the well-tested and versatile QDP++ 
framework \cite{Edwards:2004sx}, which allows for data-parallel programming on 
high performance clusters.
Unless stated differently, we will use $QL=120$ throughout this section. Furthermore, as introduced in \autoref{sec:initial_conditions}, the initial conditions in the boost invariant scenario, i.e.~the one without longitudinal fluctuations, are identical in both frameworks. We will therefore present corresponding results for the energy density solely in the expanding formulation, since the counterparts in the static box can easily be derived therefrom due to energy conservation.

\subsection{\SU{2} vs. \SU{3}}\label{sec:SU2_SU3}

\begin{figure}[!ht]
	\centerline{
	\includegraphics[scale=0.8]{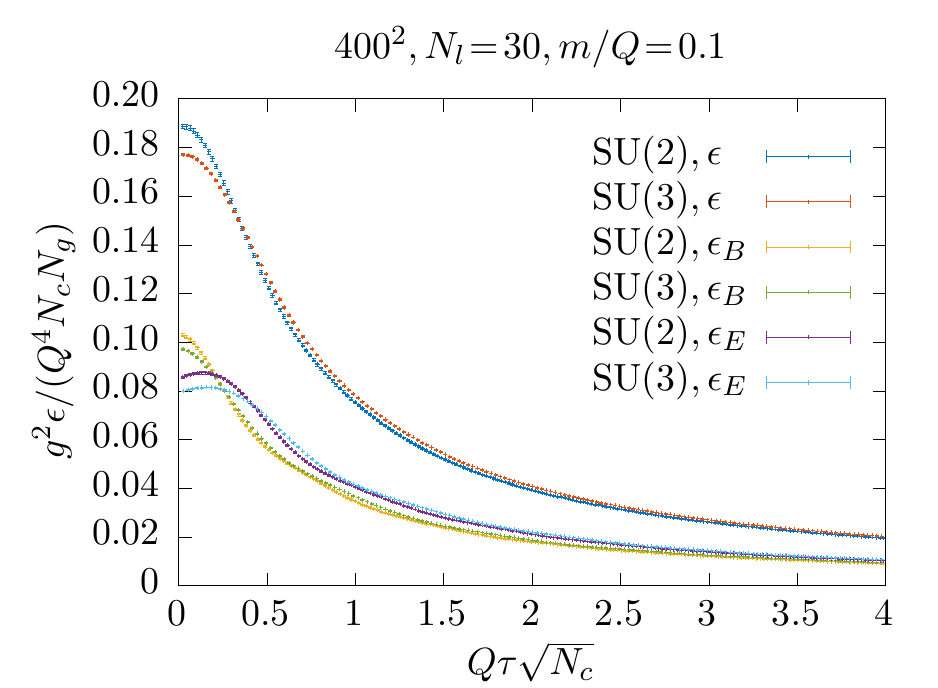}
	}
	\caption{Total energy density and its chromo-magnetic and chromo-electric components for \SU{2} and \SU{3}.}
	\label{fig:eps_Nc_dep}
\end{figure}

Performing the calculations for the realistic \SU{3} rather than \SU{2} gauge theory introduces 
roughly an additional factor of 3 in terms of computational time, 
depending on the studied observables. 
Comparing physical results between the groups is non-trivial, since the ratio $Q_s/Q$ depends 
on the number of colours, as well as our observables like the energy density. 
For the saturation scale we have $Q_s\sim\sqrt{N_c}Q$ \cite{Lappi2008} and for the initial 
energy density $g^2\epsilon(t|\tau=0) \sim N_cN_g$ \cite{Fukushima2008}. A physically meaningful,
dimensionless combination with the leading $N_c$-behaviour scaled out 
is thus $g^2\epsilon/(Q^4N_cN_g)$ plotted vs.~$\sqrt{N_c}Q\tau$. 
In \autoref{fig:eps_Nc_dep}, where we applied this rescaling\footnote{In the following, we will keep the scaling factor for the energy density, but we will drop the $\sqrt{N_c}$ normalisation factor in front of $Q\tau$ in order to ease the comparison with other works, where this is almost always neglected, too.}, we clearly see that there is no significant difference in the observables we are studying.
In particular, the sub-leading $N_c$-dependence appears to be much weaker 
than the sensitivity to the parameters of CGC initial conditions, which will be discussed 
in \autoref{sec:CGC_IC_params}. We checked this observation for several parameter settings 
with the same outcome and will therefore focus mostly on \SU{2} 
in the following, in order to reduce the numerical cost.

\subsection{Boundary effects}

\begin{figure}[t]
	\centerline{
	\hspace{0.15cm}
	\includegraphics[scale=0.8]{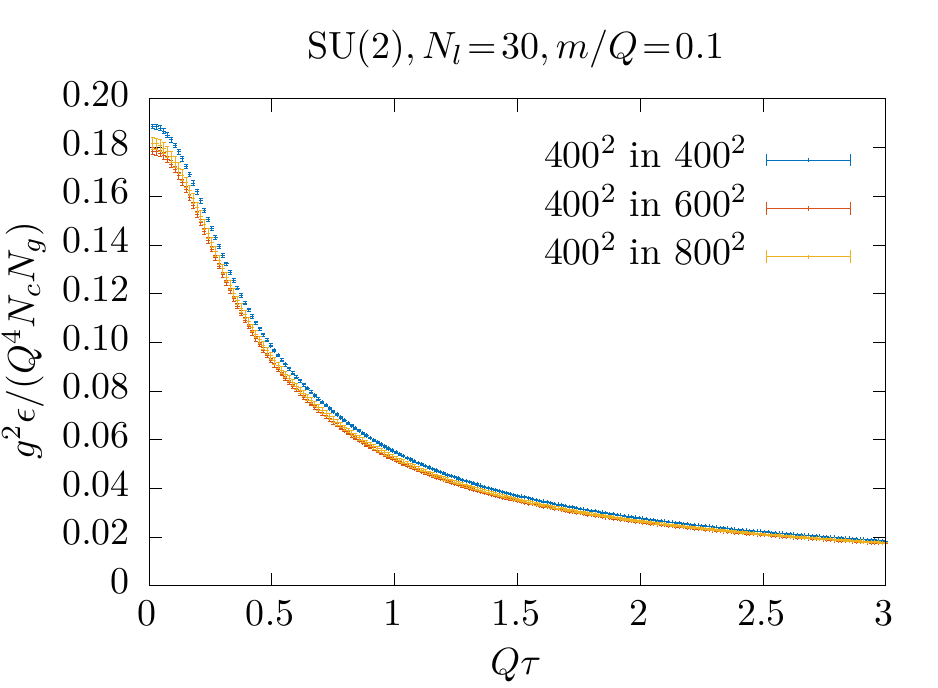}
	\includegraphics[scale=0.8]{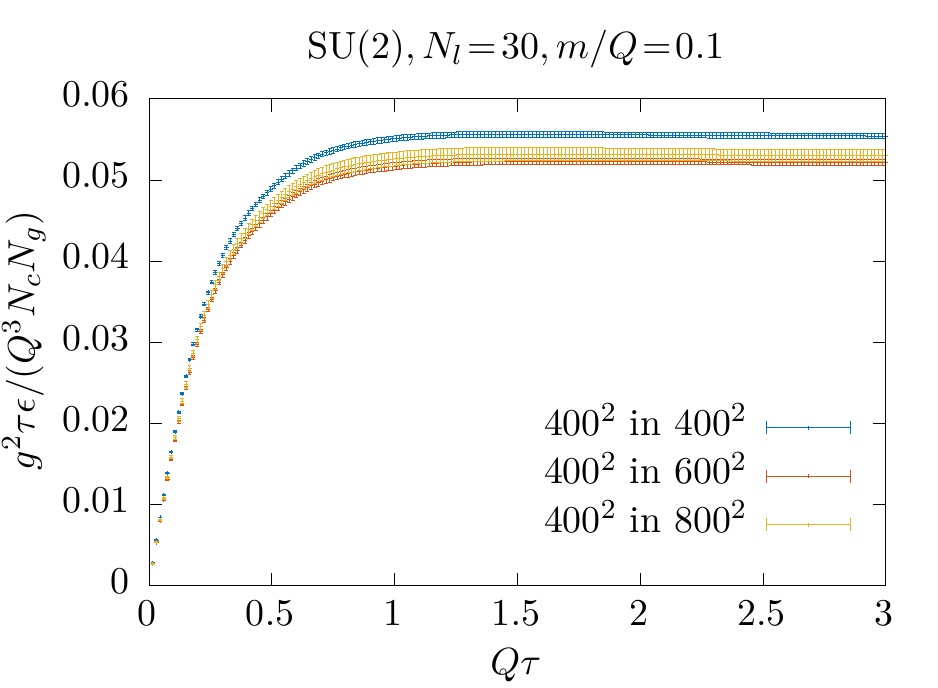}
	}
	\caption{Total energy density (left) and total energy density times the proper time (right) for a nucleus represented by $400^2$ lattice points embedded in different lattice sizes.}
	\label{fig:eps_boundary}
\end{figure}

In the MV model, the nucleus is usually "spread" over the whole lattice. This introduces 
a systematic error when using periodic boundary conditions. 
However, for our choice of parameters the total diameter of the plane representing the nucleus 
is about $\SI{12}{\femto\meter}$, which should be large enough to suppress boundary 
effects. In \autoref{fig:eps_boundary} 
we show the total energy density $\epsilon$ 
(times the proper time $\tau$) in comoving coordinates 
for three different scenarios: first, the nucleus is "spread" over the whole $400^2$ 
points on the transverse lattice plane, second, the nucleus is represented 
by $400^2$ lattice points within a $600^2$ lattice and third, the same nucleus is 
embedded in an $800^2$ lattice. We observe an effect at the 5\%-level.
We have explicitly checked that the size of finite volume effects does not change
when additional model parameters are introduced, as in the following subsections.

\subsection{Discretisation effects}

Ideally, the non-physical scales $a_\sigma$ or $a_\perp$ entering our calculations 
because of the lattice discretisation should have no effect on our results. 
On the other hand, a continuum limit does not exist for a classical theory and 
one has to investigate which values of the lattice spacing are appropriate and 
to which extent observables are affected by it.

\begin{figure}[t]
	\centerline{
	\hspace{0.1cm}
	\includegraphics[scale=0.8]{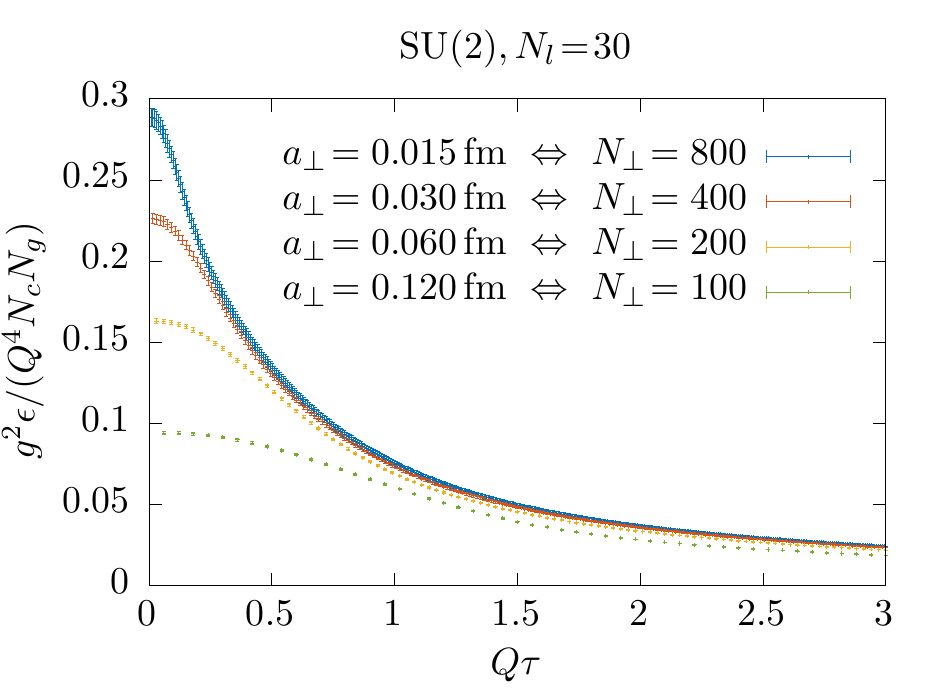}
	\includegraphics[scale=0.8]{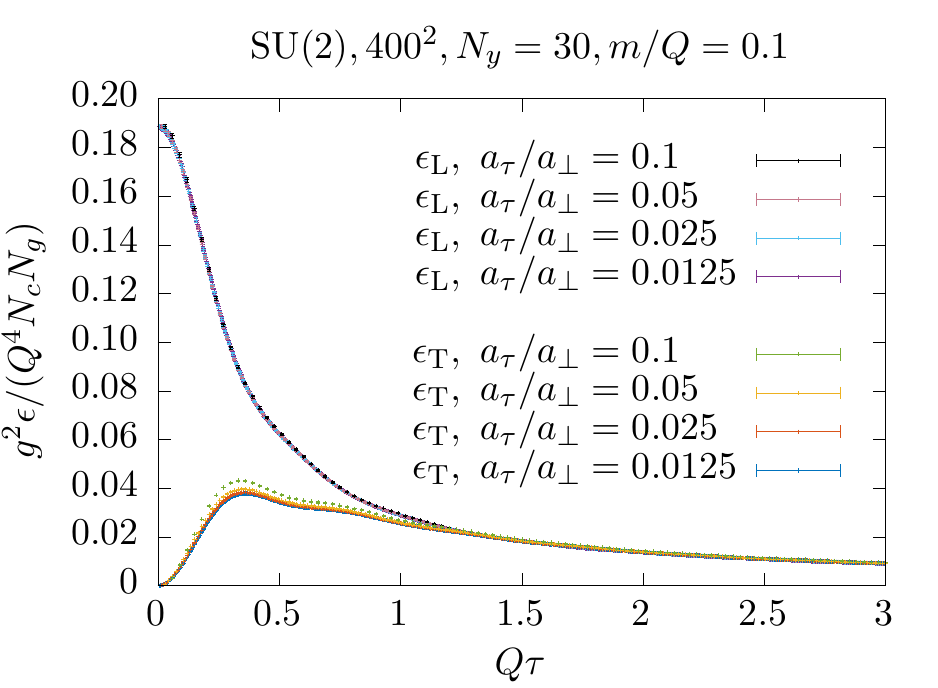}
	}
	\caption{Total energy density for different transverse lattice spacings 
$a_\perp$ (left) and different temporal lattice spacings $a_\tau$ (right).}
	\label{fig:eps_discretisation}
\end{figure}

For our problem at hand, the transverse lattice spacing is set by the number 
of lattice points spanning the size of the nucleus, cf.~\eqref{eq:a_perp}.
On a coarser lattice less momentum modes are available, which translates into lower
initial energy density for a fixed colour charge density $Q$, as shown 
in \autoref{fig:eps_discretisation} (left). 
For a non-expanding system the energy density stays
constant, thus implying large discretisation effects.  
In the expanding system, these differences are quickly diminished below percent level,
which in the literature is often interpreted as a sign for continuum-like behaviour.

Note however, that the apparent freedom to choose a lattice spacing results from
our ignorance of the detailed physics.
While yet unknown, there must be a relation $\epsilon(Q)$ between energy density
and colour charge density for given nuclei and collision energy. 
The lattice
spacing would then be fixed by matching the energy density of the classical
system to the physical one, similar to the situation in equilibrium.
 
For our further investigations we will choose a 
$400^2$ lattice, since it is a reasonable compromise between small discretisation
effects and computation time. 
As can be seen in \autoref{fig:eps_discretisation} (left), 
with this choice the discretisation effects are negligible for $Q\tau \gtrsim 0.3$. 

We also have to be sure that there are no discretisation effects coming from the numerical 
integration over the time variable. 
To this end we vary the anisotropy parameter $\xi$, with the results for the 
transverse and longitudinal energy density shown in 
\autoref{fig:eps_discretisation} (right). 
We used $\xi=20 \Leftrightarrow a_{t|\tau}=0.05\,a_{\sigma|\perp}$ for all the results 
presented in this work, since this choice leads to negligible systematic errors coming 
from our time discretisation.

\subsection{Investigation of the parameters of the CGC initial conditions}\label{sec:CGC_IC_params}

In the following we elaborate on the different parameters entering the system's description through the CGC initial conditions.

\subsubsection{Number of longitudinal sheets $N_l$}
\begin{figure}[t]
        \centerline{
        \hspace{0.1cm}
        \includegraphics[scale=0.8]{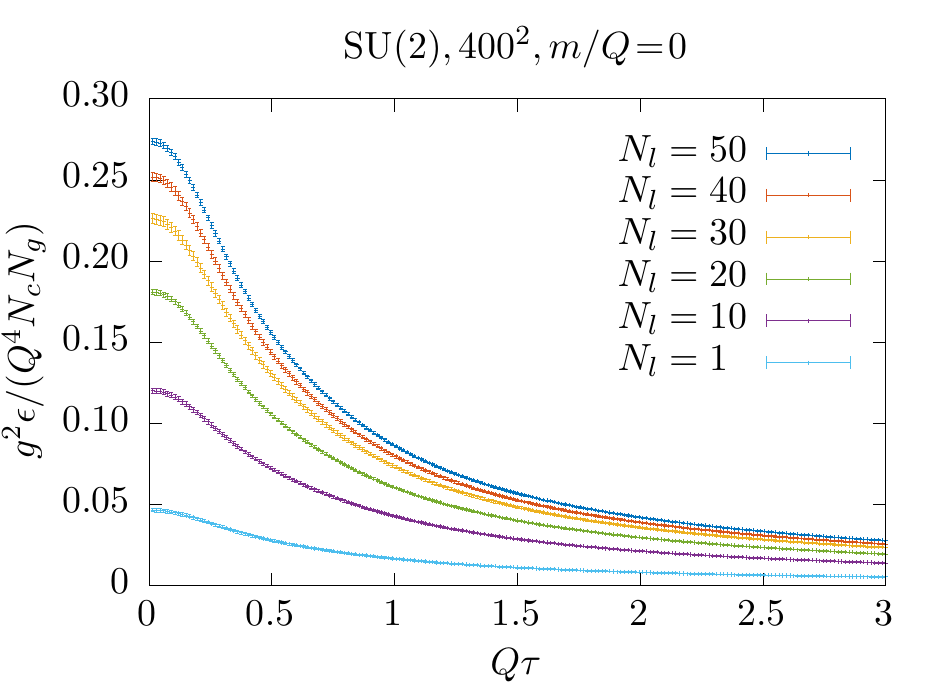}
        \includegraphics[scale=0.8]{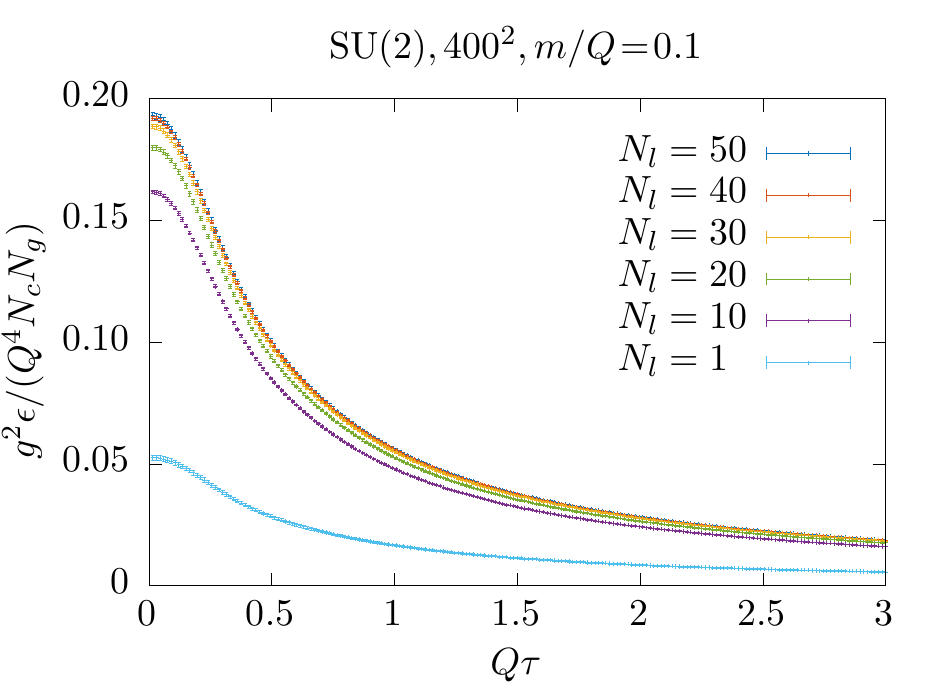}
        }
        \caption{Total energy density for different numbers of longitudinal sheets $N_l$ 
without (left) and with (right) an additional IR cutoff.}
        \label{fig:eps_Ny_dep}
\end{figure}

As shown in \cite{Fukushima2008}, the originally proposed initial conditions of the MV model
lack randomness within the longitudinal dimension. Fukushima proposed to use $N_l$ sheets of
the nucleus rather than only a single one. This is a merely technical parameter
coming from the
numerical implementation and thus vanishes in continuous time, where $N_l\!\rightarrow\!\infty$. \autoref{fig:eps_Ny_dep} shows that the total energy density depends strongly on $N_l$ for small values $\lesssim\!30$ and then saturates. This effect is amplified by adding an IR cutoff $m$, leading to a faster saturation for $m/Q=0.1$ than for $m/Q=0$. This has also been observed in \cite{Lappi2008} and can be expected: the IR cutoff introduces an additional screening of the colour sources and hence reduces the correlation length also in the rapidity direction.
The computation time of the system's initialisation grows linearly with $N_l$ and hence a reasonable choice is $N_l=30$, which we set for most of our simulations.

\subsubsection{IR cutoff $m$}
\begin{figure}[t]
	\centerline{
	\hspace{0.1cm}
	\includegraphics[scale=0.8]{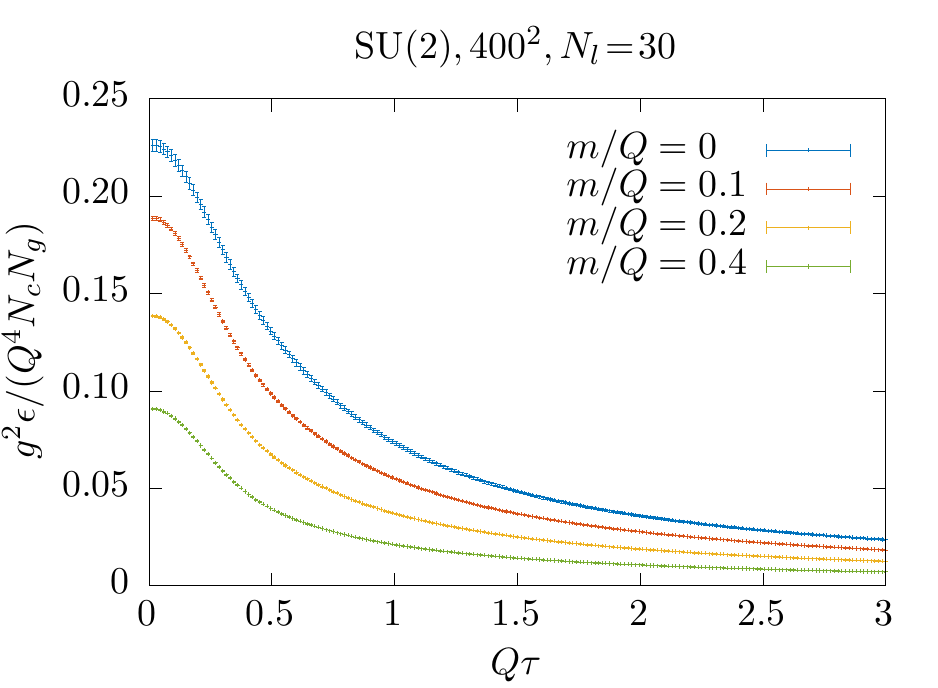}
	\includegraphics[scale=0.8]{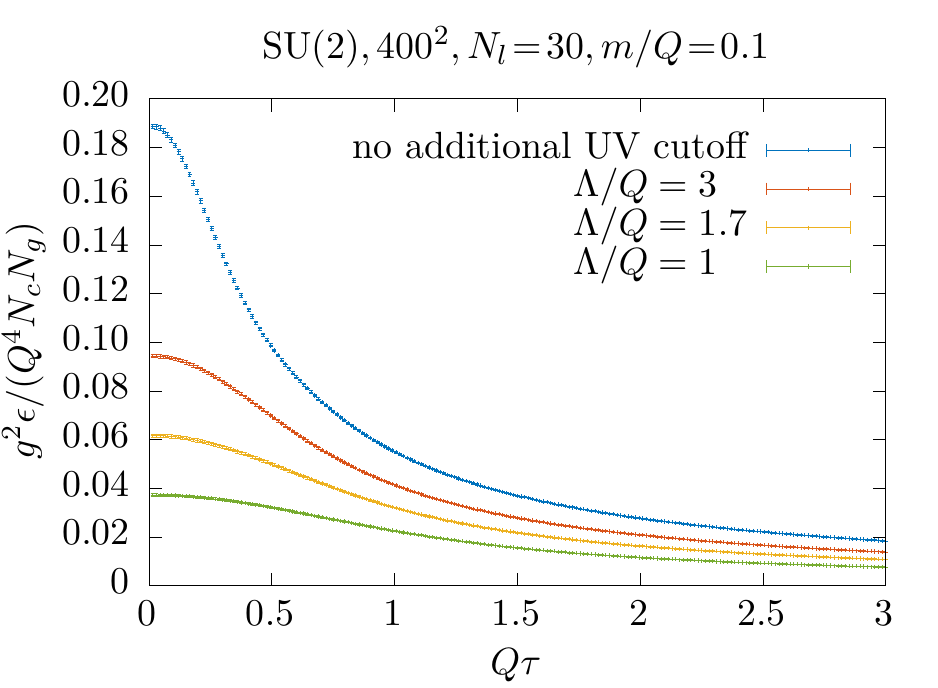}
	}
	\caption{Total energy density for different IR (left) and UV (right)
cutoff parameters.}
	\label{fig:eps_m_dep}
\end{figure}

As explained in the last section, the IR parameter $m$ provides a simple way to 
incorporate the colour neutrality phenomenon studied in \cite{Iancu2003c}. 
While $m=0.1\,Q\approx\frac{1}{R_p}$, with $R_p$ being the proton radius, is a physically
motivated choice,
the precise value of $m/Q$ has a large effect on the initial energy density which can be 
seen in \autoref{fig:eps_m_dep} (left). 
With a higher cutoff, less modes are populated to
contribute to the energy density. 
As studied in \cite{Lappi2008}, the parameter $m$ also affects the ratio $Q/Q_s$: 
at $N_l=30$ the physical saturation scale $Q_s$ is around $0.85\,Q$ for $m/Q=0.1$ 
and around $1.03\,Q$ for $m/Q=0$. Since the energy density is normalised by $Q^4$, 
this difference amounts to about a factor of 2 in the dimensionless quantity 
$\epsilon/Q_s^4$.

\begin{figure}[th]
        \centerline{
        \includegraphics[scale=0.8]{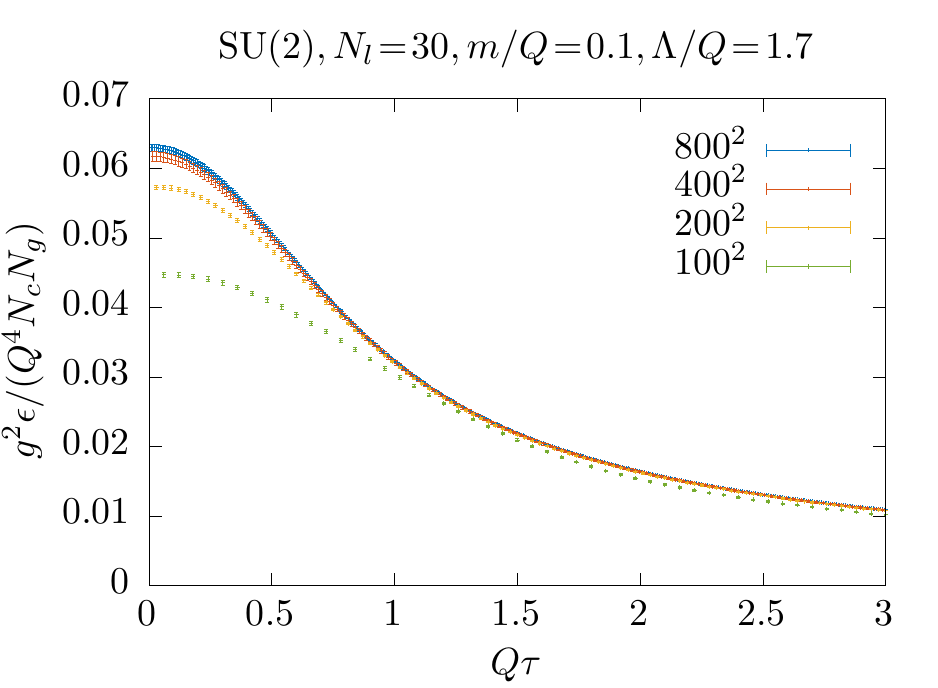}
        }
        \caption{Total energy density for different transverse lattice sizes $N_\perp^2$ with an additional UV 
cutoff of $\Lambda=1.7\,Q$.}
        \label{fig:eps_discretisation_UV_cutoff}
\end{figure}

Since the effect of $m$ is in the infrared,
it does not get washed out by the expansion of the system, 
in contrast to the discretisation effects. Hence a careful understanding 
to fix this parameter is important. For example, one might wonder whether this
inverse length scale should not also be anisotropic in the initial geometry.
In what follows we will either use $m=0$, as in the initial MV model, 
or the physically motivated choice $m/Q=0.1$.

\subsubsection{UV cutoff $\Lambda$}

As discussed in \autoref{sec:initial_conditions}, one can apply a UV cutoff $\Lambda$ 
while solving Poisson's equation \eqref{eq:Poisson}, in addition to the existing lattice 
UV cutoff. This is an additional model parameter 
limiting the initial mode population to an infrared sector determined by $\Lambda$. 
\autoref{fig:eps_m_dep} (right) shows the influence of this parameter on the 
energy density,
which gets reduced because of the missing higher modes in the Poisson equation.
This is similar to the observation we made on the IR cutoff $m$, 
but with the important difference that the ratio $Q/Q_s$ is independent 
of $\Lambda$ \cite{Fries2006}.
We are not aware of
a unique argument or procedure to set this parameter,
for the sake of comparison with the literature 
we choose $\Lambda/Q=1.7$ \cite{Fukushima:2013dma} in some of our 
later investigations. 
As a side effect, with the emphasis of
the infrared modes strengthened, the dependence of the total energy density on the lattice
spacing is reduced and 
the expanding system saturates even faster towards $a_\perp$-independent values, 
cf.~\autoref{fig:eps_discretisation_UV_cutoff} and the previous 
\autoref{fig:eps_discretisation} (left). 

\subsection{The mode spectrum}

\begin{figure}[t]
        \centerline{
        \hspace{-1cm}
        \includegraphics[scale=0.8]{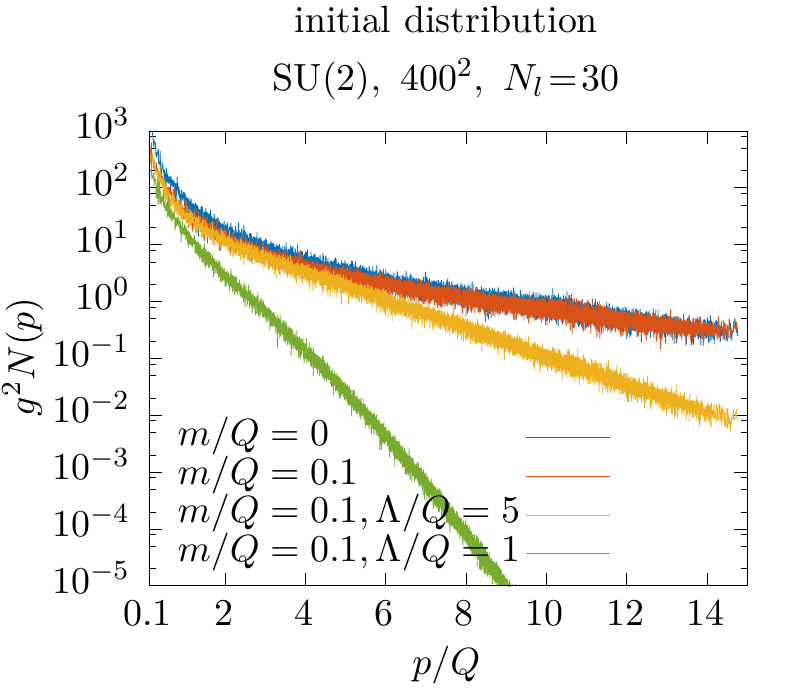}
        }
        \centerline{
        \hspace{0.1cm}
        \includegraphics[scale=0.8]{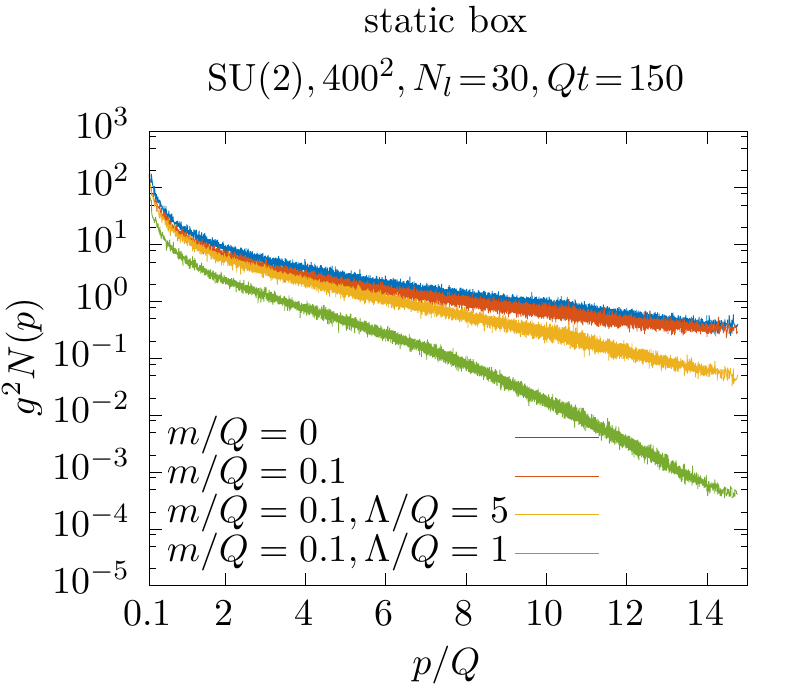}
        \includegraphics[scale=0.8]{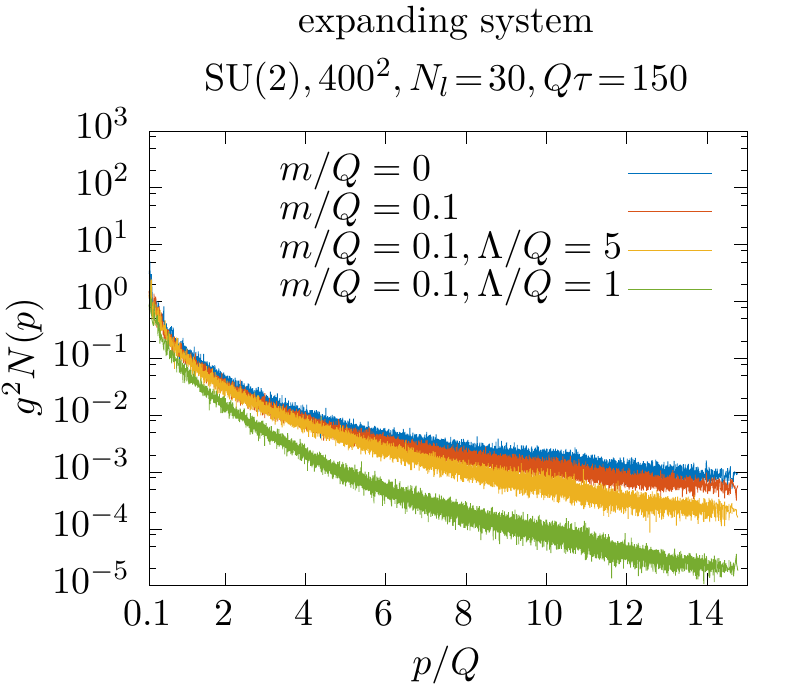}
        }
        \caption{Occupation number as a function of the momentum $p/Q$ at initial time (top)
and after the same number of
time steps in the static box (bottom left) and in the expanding formulation (bottom right). The highest momentum is defined by the lattice cutoff, $p_\mathrm{max} = \sqrt{2}\pi/a_\perp \approx 14.81\,Q$.}
        \label{fig:spec_10000}
\end{figure}

To our knowledge, the occupation number of the field modes in Fourier space is currently 
the only criterion applied
to judge the validity of the classical approximation during the time evolution
of the system.  
It is well-established that, starting from CGC initial conditions, simulations 
in a static box quickly populate higher modes, implying a breakdown of the classical 
description beyond some time. In the expanding system this process is considerably slowed down
\cite{Fukushima:2013dma,Kovchegov2005,Lappi2006b,Fujii2009}. 
We confirm these earlier findings by plotting the occupation number as a function of the momentum modes defined via \eqref{eq:modes}.

\autoref{fig:spec_10000} (top) shows the mode spectra for different model parameter 
values at initial time. In order to study the full range of the additional UV cutoff, 
we deliberately chose $\Lambda=Q$ as its smallest value, cf.~\eqref{eq:CGC_params}. 
One clearly sees that the additional UV cutoff causes a strong suppression of higher modes,
thus strengthening the validity of the classical approximation.
Another observation is that the distribution is rather independent of the IR cutoff value.
In \autoref{fig:spec_10000} (bottom) we present the evolution of the same initial configuration 
in the static and expanding framework. While without an additional UV cutoff the distributions 
nearly reach a plateau in the static box, the occupation of the higher modes 
in the expanding system stays considerably lower, thus extending the validity of the 
classical approximation. 

One can now try to get a quantitative measure of the supposed dominance of infrared modes. 
By integrating the Fourier modes of the energy density up to some
momentum scale, one can infer the energy fraction of the system contained in the modes below that scale.
For example, without applying any cutoffs, integrating modes up to $2Q\approx \SI{4}{\giga\electronvolt}$ contains 65\% of the total energy of 
the system at initial time. At $Qt|Q\tau=150$, this changes to 60\% or 77\% in the static and expanding cases, respectively.
Hence, the quality of the classical approximation deteriorates only slowly or not at all. 
Nevertheless, a significant systematic error should be expected when several 10\% of the energy is in 
the UV sector, where a running coupling and other quantum effects should be taken into account. 
This must certainly be the case when modes $\gtrsim 5Q\approx \SI{10}{\giga\electronvolt}$ get significantly populated, 
as in \autoref{fig:spec_10000}. At this stage of the evolution a better description might be obtained by
an effective kinetic theory \cite{Baier:2000sb, Arnold:2002zm, Kurkela:2016vts}, where quantum effects are already included.

\begin{figure}[t]
	\centerline{
	\hspace{0.1cm}
	\includegraphics[scale=0.8]{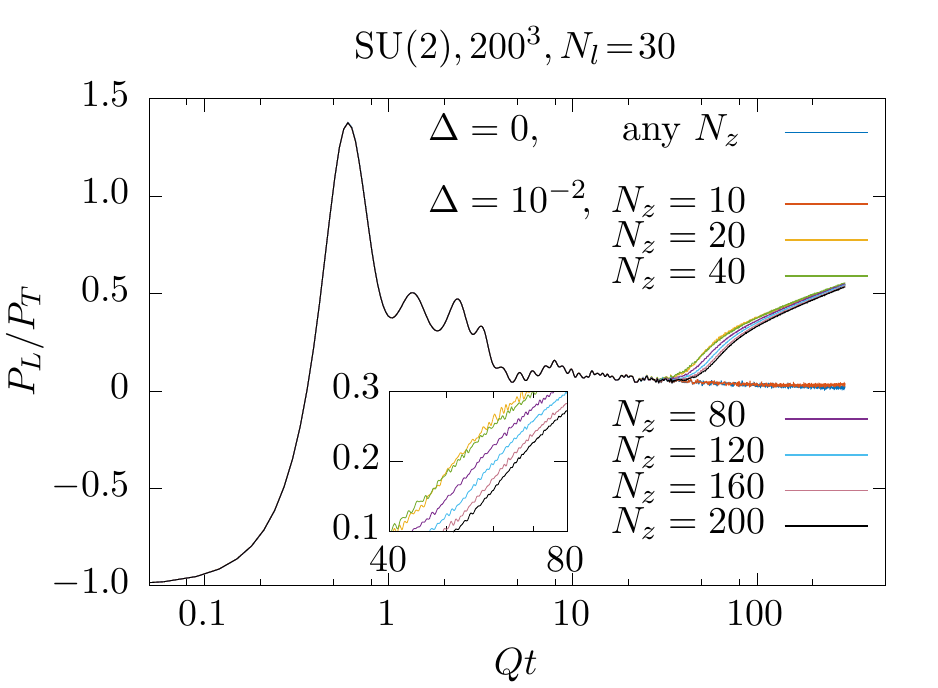}
	\includegraphics[scale=0.8]{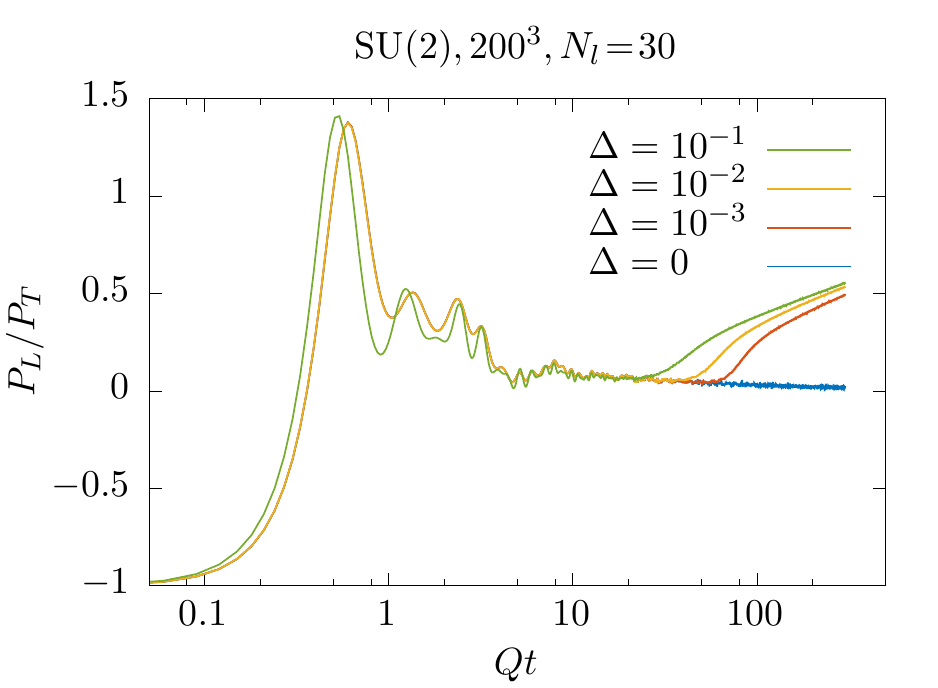}
	}
	\caption{Pressure ratio in the static box for different longitudinal lattice extents $N_z$ (left) and for different fluctuation amplitudes $\Delta$ (right).}
	\label{fig:pressure_ratio}
\end{figure}

\subsection{Isotropisation}

In this section we add small quantum fluctuations on the initial conditions,
as described by eq.~\eqref{eq:fluctuations}.
These initial fluctuations lead to an eventual
isotropisation of the system, which can be studied by the evolution of the ratio of the pressure 
components $P_L/P_T$. 
To include their effects, we have to extend our two-dimensional analysis by an 
additional longitudinal 
direction $N_{z|\eta}$, increasing the computation time linearly 
with $N_{z|\eta}$. Within our computational budget, this forces us to use smaller lattices ($200^3$)
for this section, thus inevitably increasing the cutoff and finite volume effects 
we have discussed so far. However, as we shall see, the effects of the model 
parameters are by an order of magnitude larger.

\subsubsection{Static box}

We begin with the static box. The general behaviour of the pressure ratio $P_L/P_T$ 
has been known for a while and is 
shown in \autoref{fig:pressure_ratio}. After a peak at around $Q t \approx 0.6$ 
follows an oscillating stage until the system isotropises. The oscillating stage originates 
from turbulent pattern formation and diffusion~\cite{Fukushima:2013dma,Berges:2013eia} and precludes 
a hydrodynamical description. 
We see a strong finite size effect in $N_z$, \autoref{fig:pressure_ratio} (left), 
which decreases for larger values and should vanish 
in the limit $N_z\!\rightarrow\!\infty$. For very small values of $N_z\leq 10$, 
the fluctuations cannot 
evolve and  
the system behaves as in the unperturbed $\Delta=0$ case.

\begin{table}[t]
\begin{center}
\begin{tabular}{lcccccc}
\Xhline{1pt}
&&&&&&\\[-12pt]
&& $\frac{g^2\epsilon}{Q^4N_cN_g}$ && \multicolumn{3}{c}{relative increase} \\[2pt]
&& $\Delta\!=\!0$ && $\Delta\!=\!10^{-1}$ & $\Delta\!=\!10^{-2}$ & $\Delta\!=\!10^{-3}$ \\[3pt]
no additional cutoff 									&& $0.163$ && $23.9\,\%$ & $0.239\,\%$ & $0.00239\,\%$ \\
$m/Q\!=\!0.1$ 												&& $0.122$ && $32.2\,\%$ & $0.321\,\%$ & $0.00322\,\%$ \\
$m/Q\!=\!0.1$ and $\Lambda/Q\!=\!1.7$ && $0.057$ && $68.1\,\%$ & $0.682\,\%$ & $0.00683\,\%$ \\[-12pt]
&&&&&&\\
\Xhline{1pt}
\end{tabular}
\caption{The initial total energy density and its relative increase due to the fluctuations for different cutoff setups. The statistical errors are all below the $1\,\%$-level.}
\label{tab:rel_eps_growth}
\end{center}
\end{table}

The dependence on the fluctuation amplitude $\Delta$ is studied 
in \autoref{fig:pressure_ratio} (right). In accord with expectation, 
increasing the fluctuation amplitude $\Delta$ 
reduces the isotropisation time. Note the interesting dynamics associated with this:
while for larger initial amplitudes the onset towards isotropisation 
occurs earlier, the eventual growth of the longitudinal pressure appears to be faster
for the smaller amplitudes. The initial fluctuation amplitude $\Delta$ 
also significantly affects the early behaviour of the system, causing a strong change 
of the pressure ratio and a significant increase of the energy density ($\sim\!\Delta^2$), 
as shown in \autoref{tab:rel_eps_growth}. Also the frequencies of the 
plasma oscillations are affected.
Of course,
increasing the quantum fluctuation amplitude weakens the classicality of the 
initial condition:
for $\Delta \geq 0.1$ the fluctuations already 
make up $\geq 20\%$ of the initial energy density. On the other hand,
for $\Delta \lesssim 10^{-2}$ there is no visible effect on the pressure ratio at early 
times ($Qt \lesssim 20$), and also the energy remains the same within numerical fluctuations.

\begin{table}[t]
\begin{center}
\begin{tabular}{ccccccccc}
\Xhline{1pt}
&&&&&&&&\\[-12pt]
\multicolumn{3}{c}{$200^3$ ~\&~ no additional cutoff} &&& \multicolumn{4}{c}{$200^2\!\times\!20$ ~\&~ $\Delta\!=\!10^{-2}$} \\[2pt]
\multirow{2}{*}{$\Delta\!=\!10^{-1}$} & \multirow{2}{*}{$\Delta\!=\!10^{-2}$} & \multirow{2}{*}{$\Delta\!=\!10^{-3}$} &&& no add. & \multirow{2}{*}{$\Lambda/Q\!=\!1.7$} & \multirow{2}{*}{$m/Q\!=\!0.1$} & $m/Q\!=\!0.1$ \\
&&&&& cutoff &&& $\Lambda/Q\!=\!1.7$ \\[3pt]
751 & 770 & 885 &&& 799 & 1719 & 3259 & 4736 \\[-12pt]
&&&&&&&&\\
\Xhline{1pt}
\end{tabular}
\caption{Hydrodynamisation time extrapolations in units of $Q^{-1}$ for different lattice and CGC parameter setups.}
\label{tab:hydro_times}
\end{center}
\end{table}

The hydrodynamisation time of a heavy ion collision is the time, after which hydrodynamics 
is applicable to describe the
dynamics of the system. This is commonly believed to be the case once the pressure 
ratio $P_L/P_T\geq 0.7$. For an initial
amplitude of $\Delta=10^{-2}$ and without further model cutoffs, 
this happens at $t\approx770/Q\approx\SI{76}{\femto\meter}$ in
our simulations.
This value is considerably larger than experimentally expected ones, but it is in line with 
earlier numerical results in a static box, 
e.g.~\cite{Fukushima:2013dma}.

\begin{figure}[t]
        \centerline{
        \hspace{0.1cm}
        \includegraphics[scale=0.8]{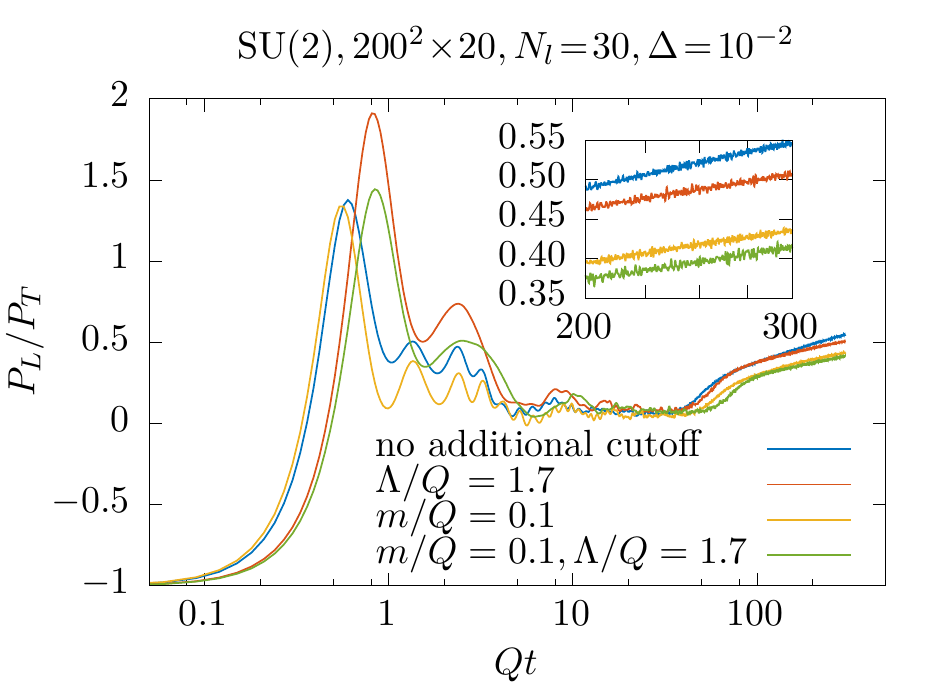}
        \includegraphics[scale=0.8]{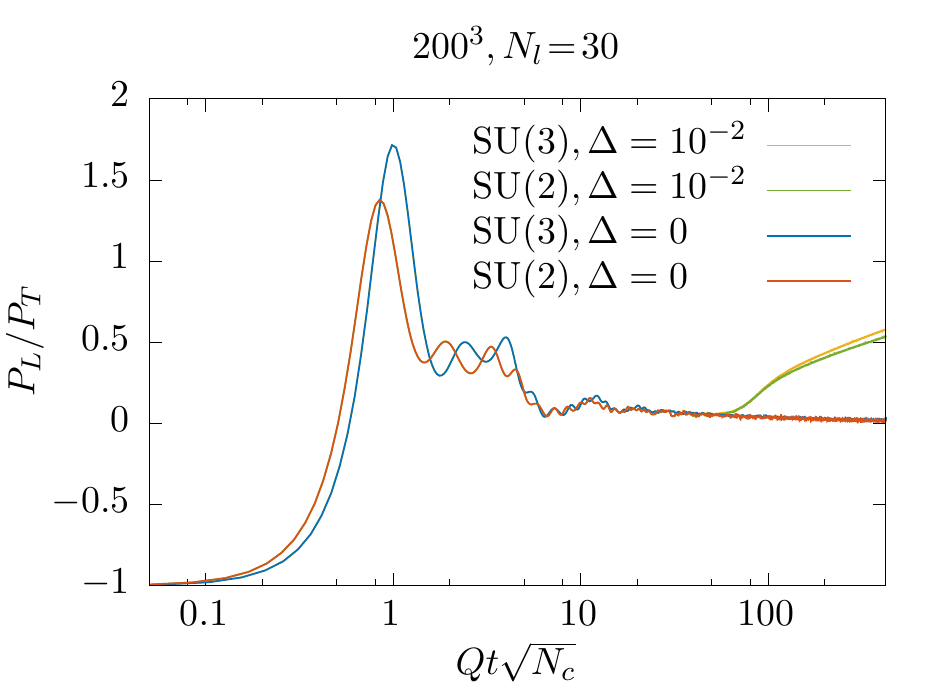}
        }
        \caption{Pressure ratio in the static box for different IR and UV cutoffs (left) and for the different
gauge groups (right).}
        \label{fig:pressure_ratio_cutoff}
\end{figure}

The pressure ratio is highly sensitive both to the additional IR and to the UV cutoff 
introduced in the initial condition, 
cf.~\autoref{fig:pressure_ratio_cutoff} (left). 
Especially 
the UV cutoff changes the qualitative shape of the curve at early times significantly. 
Furthermore, both cutoffs considerably 
slow down the process of 
isotropisation as shown in \autoref{tab:hydro_times}. The hydrodynamisation time grows by 
factors of 2-6 for cutoff values
as chosen before. Hence, a better understanding and fixing of those model parameters
is mandatory for any quantitative investigation.

In accord with \autoref{sec:SU2_SU3}, we see no significant change in the isotropisation 
time when using $N_c\!=\!3$ colours instead of $2$, 
cf.~\autoref{fig:pressure_ratio_cutoff} (right). By contrast, the details of the
oscillatory behaviour at early times differ. This implies that for the investigation 
of the properties of collective excitations as in \cite{Boguslavski:2018beu}, 
the correct gauge group will eventually be important for 
quantitative results.

\subsubsection{Chromo-Weibel instabilities}

\begin{figure}[t]
	\vspace{-45pt}
	\centerline{
	\includegraphics[scale=0.825]{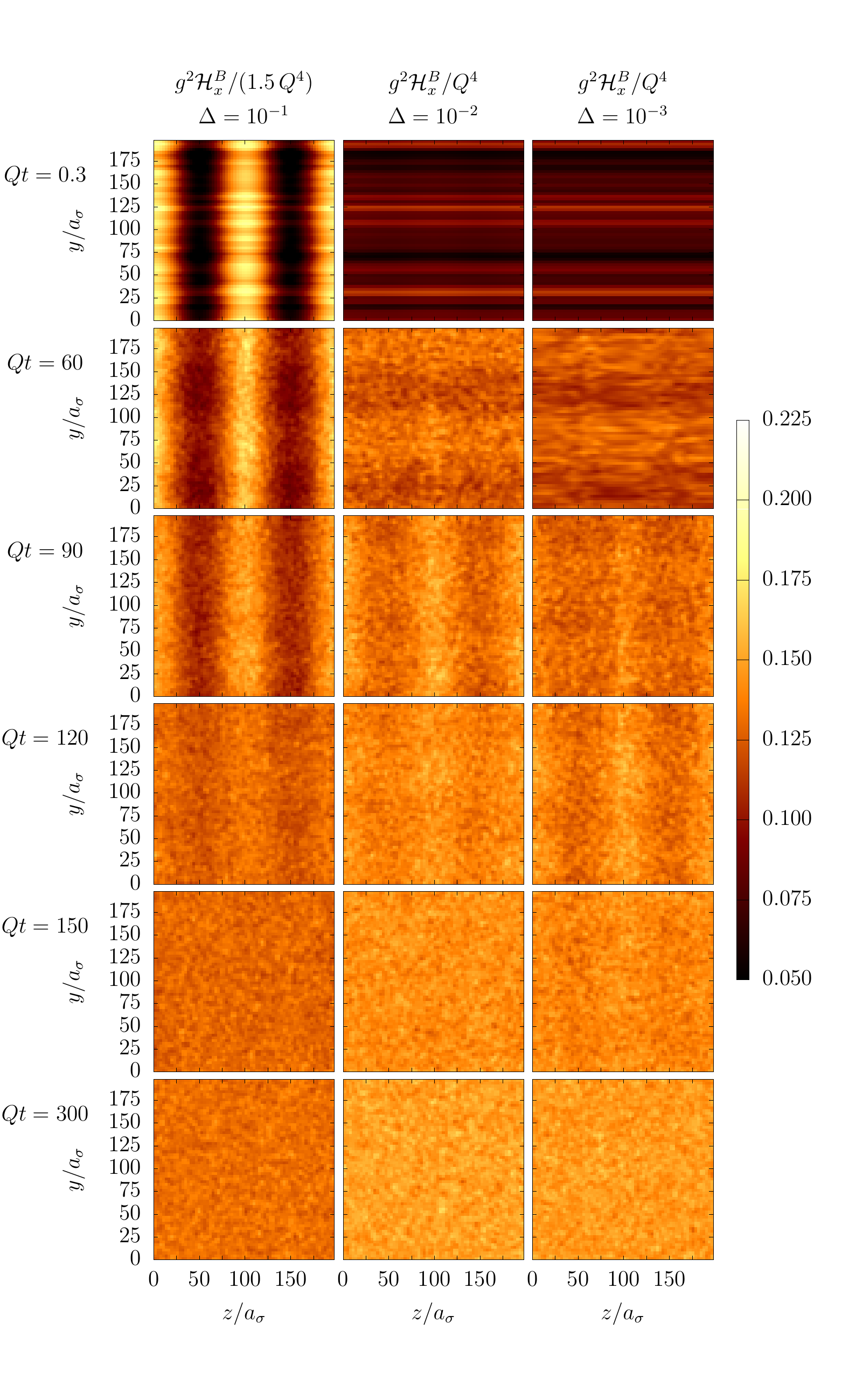}
	}
	\vspace{-25pt}
	\caption{Snapshots of the $x$-component of the chromo-magnetic energy density 
in the $yz$-plane at different times (top down: $Qt=0.3,60,90,120,150,300$) and different 
fluctuation seeds (left to right: $\Delta=10^{-1},10^{-2},10^{-3}$).}
	\label{fig:filaments}
\end{figure}

It has been suggested that the apparent rapid thermalisation during heavy ion collisions might be caused by chromo-Weibel instabilities \cite{Mrowczynski:1988dz,Pokrovsky:1988bm}. 
Indeed, the final increase of the pressure ratio towards isotropisation, as observed in 
\autoref{fig:pressure_ratio}, may be attributed to such an instability, as we now show. 
Firstly, our anisotropic initial conditions imply a fluctuating current, which is a necessary ingredient for the occurrence of a Weibel instability. 
Secondly, an instability causes a rapid population of harder modes during the evolution in time, which is clearly realised in our system, 
as shown by the occupation number in \autoref{fig:spec_10000}. 
The most striking illustration that this indeed corresponds to a chromo-Weibel instability 
is obtained by observing the 
chromo-electric and chromo-magnetic energy densities in position space, where filaments caused by the instability are clearly visible. 
\autoref{fig:filaments} shows the amplitude of the $x$-component of the chromo-magnetic energy density in the $yz$-plane while 
averaging over the remaining $x$-direction\footnote{We can of course replace $x$ by $y$ and vice versa in \autoref{fig:filaments}, since the two 
transverse directions are indistinguishable.}. Each box is a snapshot at a given time step, 
where the horizontal axis represents the longitudinal direction (along the beam line) and 
the vertical axis is in the transverse direction.

Qualitatively the snapshots do not change if we replace the chromo-magnetic energy 
density $\mathcal{H}^B$ by the chromo-electric energy density $\mathcal{H}^E$, 
with one exception: for the large fluctuation amplitude  $\Delta=10^{-1}$, the filamentation in 
$\mathcal{H}^B_{x|y}$ and $\mathcal{H}^E_z$ appears already 
at initial time, whereas it propagates 
into $\mathcal{H}^B_z$ and $\mathcal{H}^E_{x|y}$ only after a few time steps.

The pattern at $Qt = 0.3$ (first row of \autoref{fig:filaments}) for $\Delta=10^{-2}$ and 
$\Delta=10^{-3}$ represents the initial fluctuations which are independent of the 
longitudinal direction $z$. At a later time $Qt=90$ the chromo-Weibel instability is 
visible with filaments that are more pronounced for higher fluctuation seeds. 
At very late times $Qt=300$ the filaments dissolve again. Note how the detailed 
timing of the growth and decay of the filaments crucially depends on
the value of $\Delta$. It is interesting to compare these plots with 
\autoref{fig:pressure_ratio} (right): apparently the dynamical instabilities
arise late, after the oscillatory period around the onset to isotropisation. 

For consistency, we checked that indeed no filamentation arises in 
the transverse plane, as expected. 
This holds for all components of both the chromo-magnetic and for the chromo-electric 
energy density.
Instead, the average values of the energy densities are random with large fluctuations at
early stages, which get smoothed during the time evolution.
 
\subsubsection{Expanding system}

\begin{figure}[t]
        \centerline{
        \hspace{0.1cm}
        \includegraphics[scale=0.8]{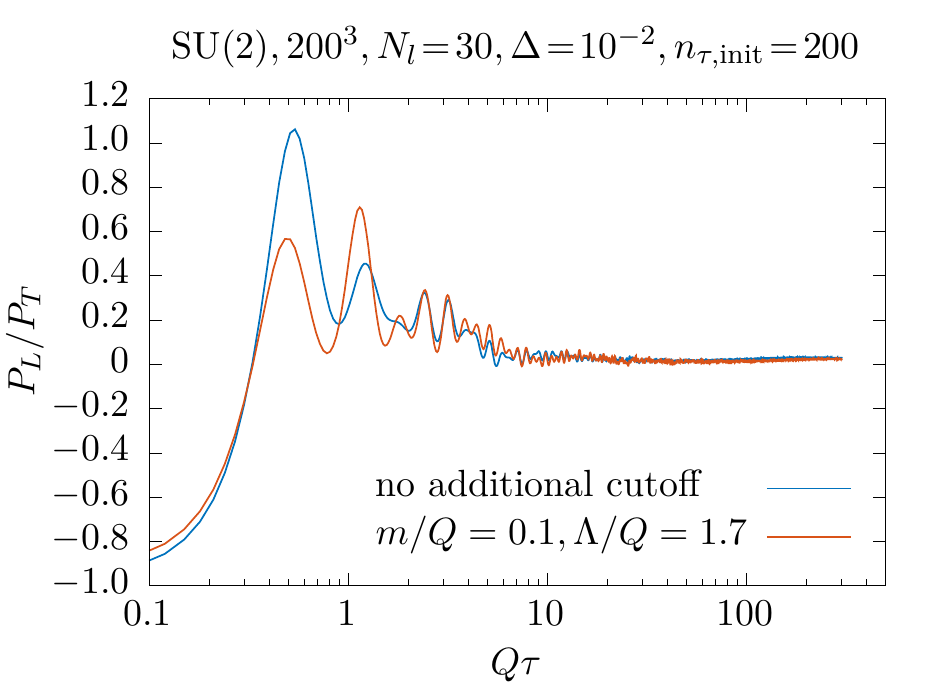}
        \includegraphics[scale=0.8]{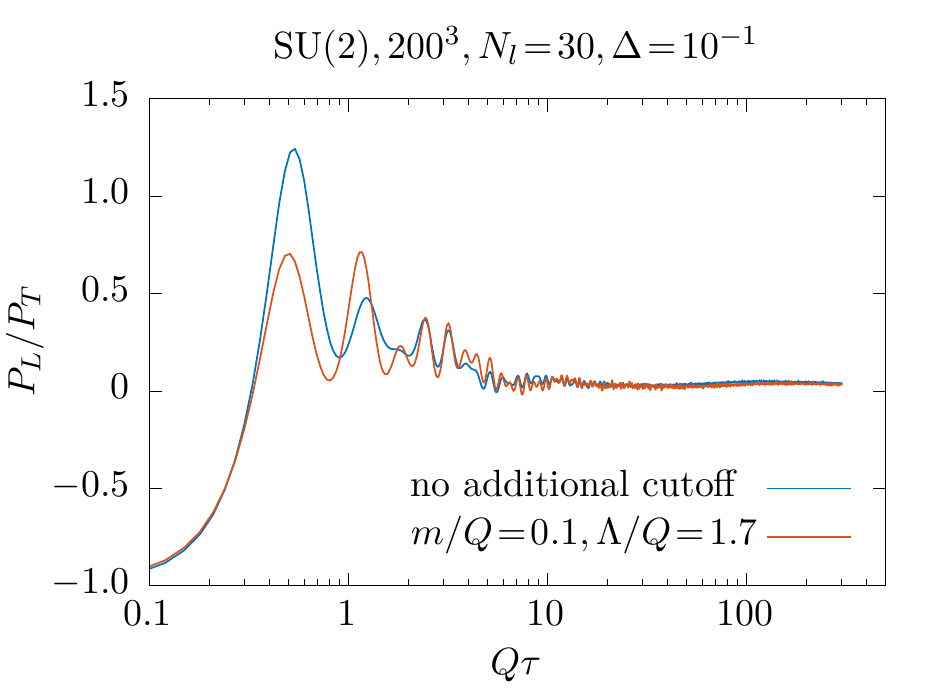}
        }
        \caption{Pressure ratio in the expanding system with different cutoffs and
fluctuation seeds. No isotropisation is observed for any choice of parameters.}
        \label{fig:pressure_ratio_Nc_dep}
\end{figure}

By contrast, in an expanding system, as realised in heavy ion collisions,
the pressure ratio does
{\it not} appear to isotropise after the oscillatory stage but settles at a small or zero value,
as shown in \autoref{fig:pressure_ratio_Nc_dep}.
This is in accord with the findings in \cite{Berges2014a} and robust under variation of all model parameters. In particular, it also holds
for the largest fluctuation seed considered, cf.~\autoref{fig:pressure_ratio_Nc_dep} (right).
Correspondingly, in the expanding system 
no dynamic filamentation takes place either. Only for fluctuation amplitudes
$\gtrsim 10^{-1}$ filaments are forced right from the beginning, 
since the initial configuration is equivalent to the one we 
have shown for the static box scenario.
The conclusion is that an expanding gluonic system dominated
by classical fields according to the CGC does not appear to isotropise and thermalise.
For future work it would now be interesting to check whether adding light quark degrees of 
freedom helps towards thermalisation, as one might expect. 

\subsection{Initial condition at fixed energy density}
\label{sec:econst}

\begin{figure}[t]
        \centerline{
        \includegraphics[scale=0.42]{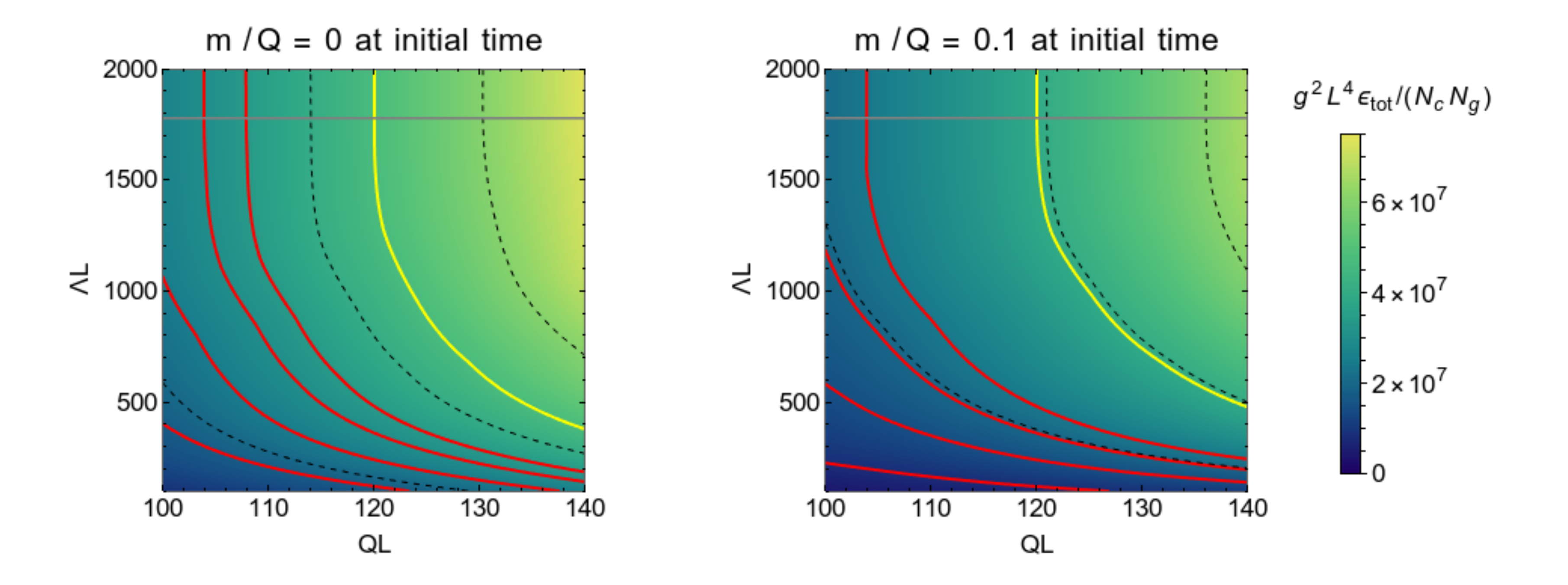}
        }
        \centerline{
        \includegraphics[scale=0.42]{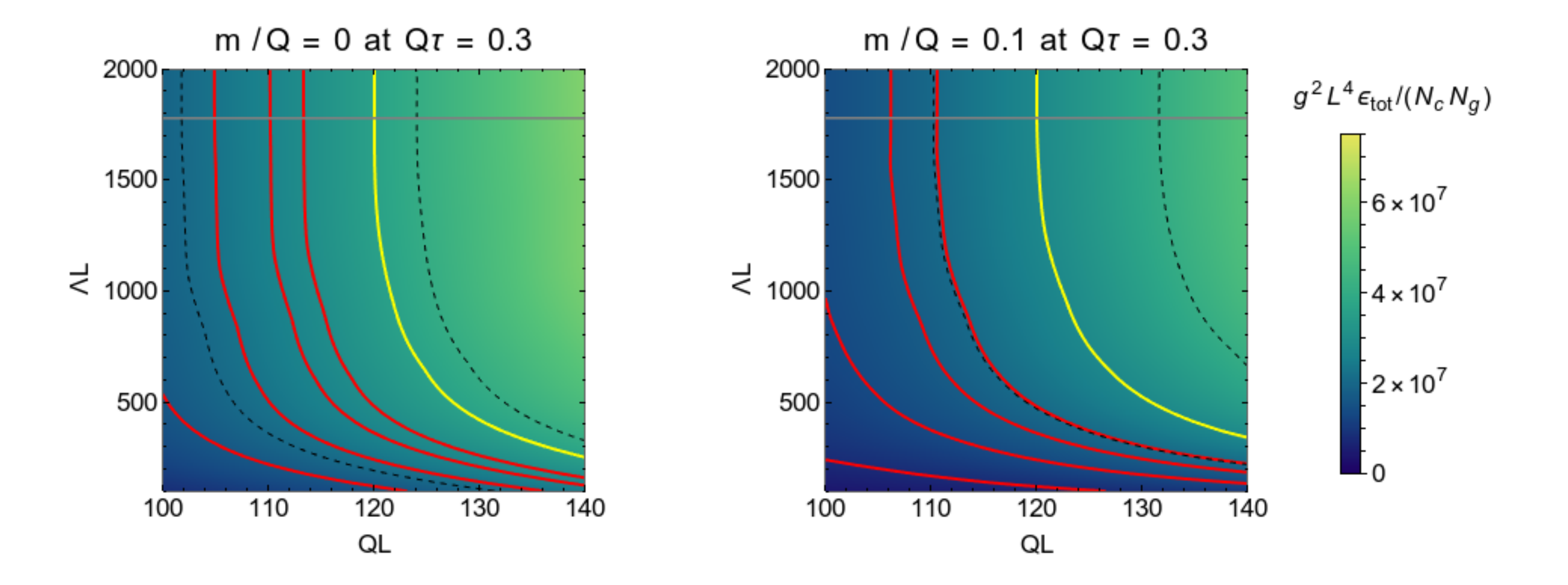}
        }
        \vspace{-5pt}
        \caption{\SU{2} total energy density on a $400^2$ lattice with $N_l\!=\!30$ as a function of $QL$ and $\Lambda L$ at initial time and at $Q\tau\!=\!0.3$. The dashed lines represent constant energy density levels at integer multiples of $10^8$. The red solid lines indicate the constant energy densities corresponding to $\Lambda\in\{Q, 2Q, 3Q,4Q\}$ at $QL\!=\!120$. The yellow solid line refers to the constant energy density contour obtained for $QL\!=\!120$ without an additional UV cutoff, which is the choice of the majority of previous studies. The gray horizontal line represents the lattice UV cutoff above which the additional UV cutoff no longer affects the system.}
        \label{fig:UV_cutoff}
\end{figure}
Altogether the numerical results of classical simulations show a large 
dependence on the various model parameters of the CGC initial condition.
This creates a difficult situation, because the initial condition and the early
stages of the evolution until freeze-out are so far not accessible experimentally. 
We now propose a different analysis of the simulation data
which should be useful in constraining model parameters such as $\Lambda, m$ and $\Delta$.

In a physical heavy ion collision the initial state is characterised by
a colour charge density, an energy density and some effective values of $\Lambda,m$ and $\Delta$.
However, these cannot all be independent, rather we must have
$\epsilon=\epsilon(Q,\Lambda,\ldots)$, where the detailed relation is fixed by the type
of nuclei and their collision energy.
We should thus analyse computations with fixed initial energy density
$L^4\epsilon$, while varying the model parameters. The outcome of such an
investigation for $\Lambda$ and $Q$ are
the contour plots shown in \autoref{fig:UV_cutoff}. We consider $Q\tau=0.3$
as well, since then even without an additional UV cutoff the discretisation effects are
negligible for $N_\perp=400$, cf.~\autoref{fig:eps_discretisation}.
In the same Figure we also compare the situation with an additional IR cutoff as
discussed earlier. Thus, to the extent that the energy density as a function of time 
can be determined
experimentally, it should be possible to establish relations between the parameters
$Q,\Lambda$ and $m$ to further constrain the initial state.

\begin{figure}[t]
        \centerline{
        \hspace{0.1cm}
        \includegraphics[scale=0.5]{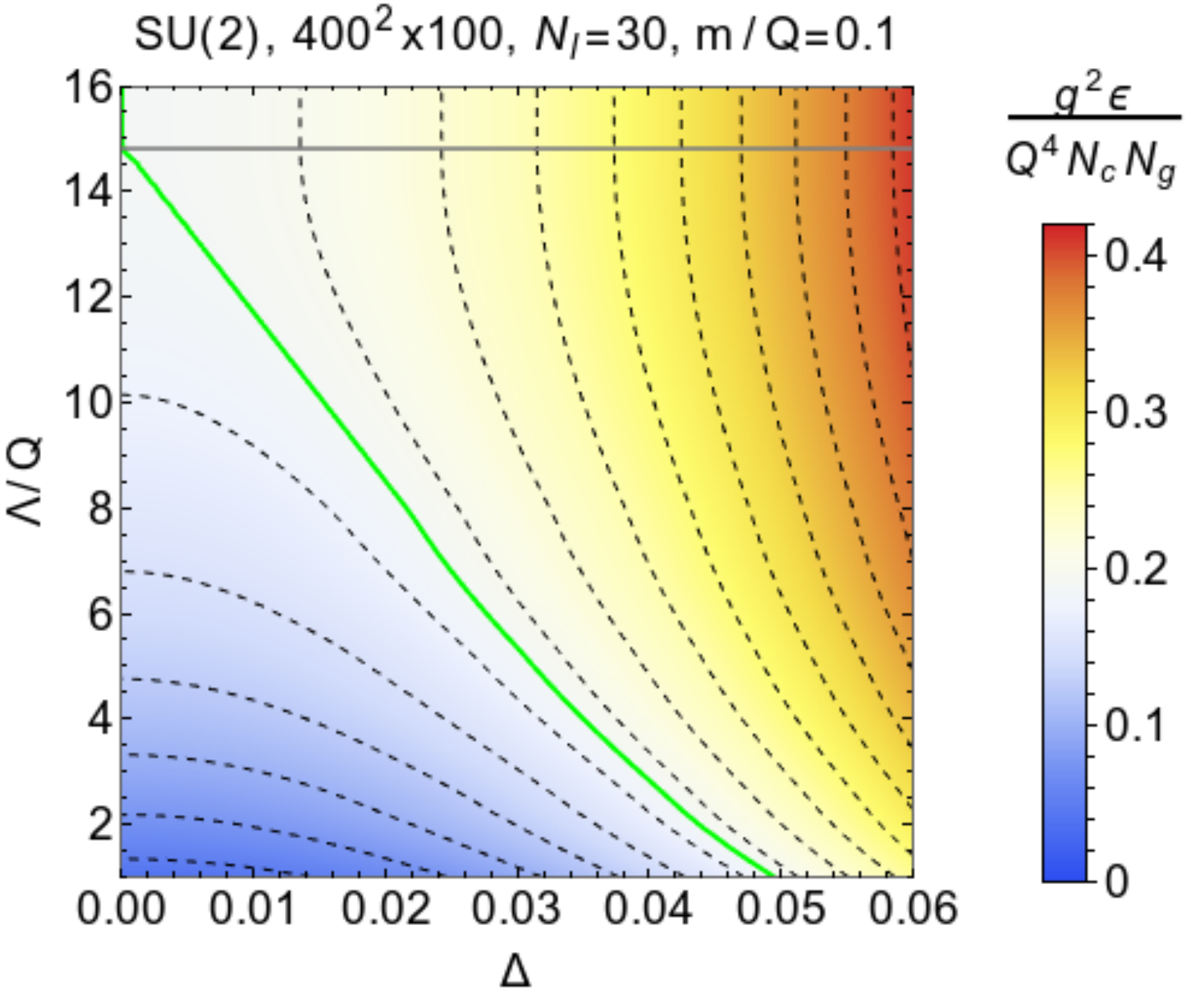}
        }
        \caption{Total energy density for different fluctuation seeds $\Delta$ and different UV cutoffs $\Lambda/Q$ at initial time. The dashed lines represent constant energy density levels in
multiples of $0.025$. The solid green line corresponds to the energy density obtained without an additional UV cutoff and no fluctuations, i.e.~$\Delta\!=\!0$. The horizontal grey line represents the lattice
cutoff above which the additional cutoff $\Lambda$ no longer affects the system.}
        \label{fig:Delta_Lambda}
\end{figure}

The same consideration can be applied to study the fluctuation amplitude.
\autoref{fig:Delta_Lambda} shows contours of
fixed energy density $\epsilon/Q^4$ in the
$(\Delta,\Lambda/Q)$ plane, where $\Delta=0$ represents the classical MV initial conditions, 
i.e., the tree-level CGC description
without any quantum fluctuations, and we have chosen $QL=120$. 
Clearly, similar studies
can be made for any pairing of the model parameters 
at any desired time during the evolution and should help in establishing 
relations between them in order to constrain the initial conditions.

\section{Conclusions}\label{sec:conclusion}

We presented a systematic investigation of the dependence of 
the energy density and the pressure on the parameters entering the lattice description 
of 
classical Yang-Mills theory, starting from the CGC initial conditions. 
This was done in a static box framework as well as in an expanding geometry and both 
for $N_c=2$ and $N_c=3$ colours. 

After the leading $N_c$-dependence is factored out, deviations between the 
\SU{2} and the 
\SU{3} formulation are small and only visible in the details of the evolution 
during the early turbulent stage.
This is not surprising in a classical treatment, 
since in the language of Feynman diagrams most of
the subleading $N_c$-behaviour is contained in loop, i.e.~quantum, corrections. 

Finite volume effects are related to the treatment of the boundary of the colliding
nuclei and their embedding on the lattice. Given 
sizes of $\sim\!\SI{10}{\femto\meter}$, such effects are at a mild 5\%-level. 
Note, however, that this 
effect is larger than the finite size effects 
of the same box on the vacuum hadron spectrum, as expected for a many-particle problem.

The choice of the lattice spacing affects the number of modes available in the 
field theory and
thus significantly influences the relation between the initial colour distribution 
and the 
total energy of the system. In the static box, 
all further evolution is naturally affected by this.
Since the classical theory has no continuum limit, 
the lattice spacing would need to be fixed
by some matching condition at the initial stage. 
By contrast, in the expanding system the energy density quickly diminishes and 
the effect of the lattice spacing is washed out. 

A quantitatively much larger and significant 
role is played by the model parameters of the initial 
conditions, 
specifically additional IR and UV cutoffs affecting the distribution of modes and the 
amplitude of initial quantum fluctuations, whose presence is a necessary condition for 
isotropisation. 
For the static box we presented direct evidence for  
isotropisation to proceed through the emergence of chromo-Weibel instabilities,
which are clearly visible as filamentation of the energy density. 
However, the hydrodynamisation time is unphysically large and gets increased further
by additional IR- and UV-cutoffs in the initial condition. Without quantitative 
knowledge of these parameters, the hydrodynamisation time varies within
a factor of five. 
We suggested a method to study the parameters' influence on the system at 
constant initial energy densities. This allows to establish
relations between different parameter sets that should be useful to constrain their 
values.

Rather strikingly, no combination of model parameters
leads to isotropisation in the expanding classical gluonic system.

\section*{ACKNOWLEDGMENTS}

We thank K.~Fukushima, J.~Glesaaen, M.~Greif, H.~van Hees, A.~Mazeliauskas, B.~Schenke, J.~Scheunert, S.~Schlichting and R.~Venugopalan for useful discussions. We are grateful to M.~Attems and C.~Schaefer for collaboration during the initial stages of this project and to FUCHS- and LOEWE-CSC high-performance computers of the Frankfurt University for providing computational resources. O.P. and B.W. are supported by the Helmholtz International Center for FAIR within the framework of the LOEWE program launched by the State of Hesse. S.Z. acknowledges support by the DFG Collaborative Research Centre SFB 1225 (ISOQUANT).

\bibliographystyle{utphys}
\bibliography{./literature}

\providecommand{\href}[2]{#2}\begingroup\raggedright\begin{thebibliography}{10}

\bibitem{Heinz:2001xi}
U.~W. Heinz and P.~F. Kolb, ``{Early thermalization at RHIC},''
  \href{http://dx.doi.org/10.1016/S0375-9474(02)00714-5}{{\em Nucl.Phys.}
  {\bfseries A702} (2002) 269--280},
\href{http://arxiv.org/abs/hep-ph/0111075}{{\ttfamily arXiv:hep-ph/0111075
  [hep-ph]}}.

\bibitem{Romatschke:2007mq}
P.~Romatschke and U.~Romatschke, ``{Viscosity Information from Relativistic
  Nuclear Collisions: How Perfect is the Fluid Observed at RHIC?},''
  \href{http://dx.doi.org/10.1103/PhysRevLett.99.172301}{{\em Phys.Rev.Lett.}
  {\bfseries 99} (2007) 172301},
\href{http://arxiv.org/abs/0706.1522}{{\ttfamily arXiv:0706.1522 [nucl-th]}}.

\bibitem{Iancu2001}
E.~Iancu, A.~Leonidov, and L.~D. McLerran, ``{Nonlinear gluon evolution in the
  color glass condensate. 1.},''
  \href{http://dx.doi.org/10.1016/S0375-9474(01)00642-X}{{\em Nucl. Phys.}
  {\bfseries A692} (2001) 583--645},
\href{http://arxiv.org/abs/hep-ph/0011241}{{\ttfamily arXiv:hep-ph/0011241
  [hep-ph]}}.

\bibitem{Kolb2000}
P.~F. Kolb, J.~Sollfrank, and U.~W. Heinz, ``{Anisotropic transverse flow and
  the quark hadron phase transition},''
  \href{http://dx.doi.org/10.1103/PhysRevC.62.054909}{{\em Phys. Rev.}
  {\bfseries C62} (2000) 054909},
\href{http://arxiv.org/abs/hep-ph/0006129}{{\ttfamily arXiv:hep-ph/0006129
  [hep-ph]}}.

\bibitem{Heinz:1985vf}
U.~W. Heinz, ``{QUARK - GLUON TRANSPORT THEORY},''
\href{http://dx.doi.org/10.1016/0375-9474(84)90579-7}{{\em Nucl.Phys.}
  {\bfseries A418} (1984) 603C--612C}.

\bibitem{Mrowczynski:1988dz}
S.~Mrowczynski, ``{Stream Instabilities of the Quark - Gluon Plasma},''
\href{http://dx.doi.org/10.1016/0370-2693(88)90124-4,
  10.1016/j.physletb.2007.09.039}{{\em Phys.Lett.} {\bfseries B214} (1988)
  587}.

\bibitem{Pokrovsky:1988bm}
Y.~Pokrovsky and A.~Selikhov, ``{Filamentation in a Quark - Gluon Plasma},''
{\em JETP Lett.} {\bfseries 47} (1988) 12--14.

\bibitem{Mrowczynski:1993qm}
S.~Mrowczynski, ``{Plasma instability at the initial stage of ultrarelativistic
  heavy ion collisions},''
\href{http://dx.doi.org/10.1016/0370-2693(93)91330-P}{{\em Phys.Lett.}
  {\bfseries B314} (1993) 118--121}.

\bibitem{Blaizot:2001nr}
J.-P. Blaizot and E.~Iancu, ``{The Quark gluon plasma: Collective dynamics and
  hard thermal loops},''
  \href{http://dx.doi.org/10.1016/S0370-1573(01)00061-8}{{\em Phys.Rept.}
  {\bfseries 359} (2002) 355--528},
\href{http://arxiv.org/abs/hep-ph/0101103}{{\ttfamily arXiv:hep-ph/0101103
  [hep-ph]}}.

\bibitem{Romatschke:2003ms}
P.~Romatschke and M.~Strickland, ``{Collective modes of an anisotropic quark
  gluon plasma},'' \href{http://dx.doi.org/10.1103/PhysRevD.68.036004}{{\em
  Phys.Rev.} {\bfseries D68} (2003) 036004},
\href{http://arxiv.org/abs/hep-ph/0304092}{{\ttfamily arXiv:hep-ph/0304092
  [hep-ph]}}.

\bibitem{Arnold:2003rq}
P.~B. Arnold, J.~Lenaghan, and G.~D. Moore, ``{QCD plasma instabilities and
  bottom up thermalization},''
  \href{http://dx.doi.org/10.1088/1126-6708/2003/08/002}{{\em JHEP} {\bfseries
  0308} (2003) 002},
\href{http://arxiv.org/abs/hep-ph/0307325}{{\ttfamily arXiv:hep-ph/0307325
  [hep-ph]}}.

\bibitem{Carrington:2014bla}
M.~E. Carrington, K.~Deja, and S.~Mrowczynski, ``{Plasmons in Anisotropic
  Quark-Gluon Plasma},''
  \href{http://dx.doi.org/10.1103/PhysRevC.90.034913}{{\em Phys.Rev.}
  {\bfseries C90} (2014) 034913},
\href{http://arxiv.org/abs/1407.2764}{{\ttfamily arXiv:1407.2764 [hep-ph]}}.

\bibitem{Romatschke:2005pm}
P.~Romatschke and R.~Venugopalan, ``{Collective non-Abelian instabilities in a
  melting color glass condensate},''
  \href{http://dx.doi.org/10.1103/PhysRevLett.96.062302}{{\em Phys.Rev.Lett.}
  {\bfseries 96} (2006) 062302},
\href{http://arxiv.org/abs/hep-ph/0510121}{{\ttfamily arXiv:hep-ph/0510121
  [hep-ph]}}.

\bibitem{Romatschke2006b}
P.~Romatschke and R.~Venugopalan, ``{The Unstable Glasma},''
  \href{http://dx.doi.org/10.1103/PhysRevD.74.045011}{{\em Phys. Rev.}
  {\bfseries D74} (2006) 045011},
\href{http://arxiv.org/abs/hep-ph/0605045}{{\ttfamily arXiv:hep-ph/0605045
  [hep-ph]}}.

\bibitem{Fukushima2011}
K.~Fukushima, ``{Evolving Glasma and Kolmogorov Spectrum},''
  \href{http://dx.doi.org/10.5506/APhysPolB.42.2697}{{\em Acta Phys. Polon.}
  {\bfseries B42} (2011) 2697--2715},
\href{http://arxiv.org/abs/1111.1025}{{\ttfamily arXiv:1111.1025 [hep-ph]}}.

\bibitem{Berges:2012iw}
J.~Berges, K.~Boguslavski, and S.~Schlichting, ``{Nonlinear amplification of
  instabilities with longitudinal expansion},''
  \href{http://dx.doi.org/10.1103/PhysRevD.85.076005}{{\em Phys.Rev.}
  {\bfseries D85} (2012) 076005},
\href{http://arxiv.org/abs/1201.3582}{{\ttfamily arXiv:1201.3582 [hep-ph]}}.

\bibitem{Berges:2013eia}
J.~Berges, K.~Boguslavski, S.~Schlichting, and R.~Venugopalan, ``{Turbulent
  thermalization process in heavy-ion collisions at ultrarelativistic
  energies},'' \href{http://dx.doi.org/10.1103/PhysRevD.89.074011}{{\em
  Phys.Rev.} {\bfseries D89} (2014) 074011},
\href{http://arxiv.org/abs/1303.5650}{{\ttfamily arXiv:1303.5650 [hep-ph]}}.

\bibitem{Fukushima:2013dma}
K.~Fukushima, ``{Turbulent pattern formation and diffusion in the early-time
  dynamics in relativistic heavy-ion collisions},''
  \href{http://dx.doi.org/10.1103/PhysRevC.89.024907}{{\em Phys.Rev.}
  {\bfseries C89} no.~2, (2014) 024907},
\href{http://arxiv.org/abs/1307.1046}{{\ttfamily arXiv:1307.1046 [hep-ph]}}.

\bibitem{Gelis2013a}
F.~Gelis, ``{Color Glass Condensate and Glasma},''
  \href{http://dx.doi.org/10.1142/S0217751X13300019}{{\em Int. J. Mod. Phys.}
  {\bfseries A28} (2013) 1330001},
\href{http://arxiv.org/abs/1211.3327}{{\ttfamily arXiv:1211.3327 [hep-ph]}}.

\bibitem{Kurkela:2011ti}
A.~Kurkela and G.~D. Moore, ``{Thermalization in Weakly Coupled Nonabelian
  Plasmas},'' \href{http://dx.doi.org/10.1007/JHEP12(2011)044}{{\em JHEP}
  {\bfseries 1112} (2011) 044},
\href{http://arxiv.org/abs/1107.5050}{{\ttfamily arXiv:1107.5050 [hep-ph]}}.

\bibitem{Kurkela:2011ub}
A.~Kurkela and G.~D. Moore, ``{Bjorken Flow, Plasma Instabilities, and
  Thermalization},'' \href{http://dx.doi.org/10.1007/JHEP11(2011)120}{{\em
  JHEP} {\bfseries 1111} (2011) 120},
\href{http://arxiv.org/abs/1108.4684}{{\ttfamily arXiv:1108.4684 [hep-ph]}}.

\bibitem{Romatschke:2006wg}
P.~Romatschke and A.~Rebhan, ``{Plasma Instabilities in an Anisotropically
  Expanding Geometry},''
  \href{http://dx.doi.org/10.1103/PhysRevLett.97.252301}{{\em Phys.Rev.Lett.}
  {\bfseries 97} (2006) 252301},
\href{http://arxiv.org/abs/hep-ph/0605064}{{\ttfamily arXiv:hep-ph/0605064
  [hep-ph]}}.

\bibitem{Rebhan:2008uj}
A.~Rebhan, M.~Strickland, and M.~Attems, ``{Instabilities of an anisotropically
  expanding non-Abelian plasma: 1D+3V discretized hard-loop simulations},''
  \href{http://dx.doi.org/10.1103/PhysRevD.78.045023}{{\em Phys.Rev.}
  {\bfseries D78} (2008) 045023},
\href{http://arxiv.org/abs/0802.1714}{{\ttfamily arXiv:0802.1714 [hep-ph]}}.

\bibitem{Attems:2012js}
M.~Attems, A.~Rebhan, and M.~Strickland, ``{Instabilities of an anisotropically
  expanding non-Abelian plasma: 3D+3V discretized hard-loop simulations},''
  \href{http://dx.doi.org/10.1103/PhysRevD.87.025010}{{\em Phys.Rev.}
  {\bfseries D87} (2013) 025010},
\href{http://arxiv.org/abs/1207.5795}{{\ttfamily arXiv:1207.5795 [hep-ph]}}.

\bibitem{Attems2016}
M.~Attems, O.~Philipsen, C.~Sch\"afer, B.~Wagenbach, and S.~Zafeiropoulos, ``{A
  real-time lattice simulation of the thermalization of a gluon plasma: first
  results},'' \href{http://dx.doi.org/10.5506/APhysPolBSupp.9.603}{{\em Acta
  Phys. Polon. Supp.} {\bfseries 9} (2016) 603},
\href{http://arxiv.org/abs/1605.07064}{{\ttfamily arXiv:1605.07064 [hep-ph]}}.

\bibitem{McLerran2002}
L.~D. McLerran, ``{The Color glass condensate and small x physics: Four
  lectures},'' \href{http://dx.doi.org/10.1007/3-540-45792-5_8}{{\em Lect.
  Notes Phys.} {\bfseries 583} (2002) 291--334},
\href{http://arxiv.org/abs/hep-ph/0104285}{{\ttfamily arXiv:hep-ph/0104285
  [hep-ph]}}.

\bibitem{Iancu:2003xm}
E.~Iancu and R.~Venugopalan, ``{The Color glass condensate and high-energy
  scattering in QCD},''
\href{http://arxiv.org/abs/hep-ph/0303204}{{\ttfamily arXiv:hep-ph/0303204
  [hep-ph]}}.

\bibitem{McLerran:1993ka}
L.~D. McLerran and R.~Venugopalan, ``{Gluon distribution functions for very
  large nuclei at small transverse momentum},''
  \href{http://dx.doi.org/10.1103/PhysRevD.49.3352}{{\em Phys.Rev.} {\bfseries
  D49} (1994) 3352--3355},
\href{http://arxiv.org/abs/hep-ph/9311205}{{\ttfamily arXiv:hep-ph/9311205
  [hep-ph]}}.

\bibitem{Lappi2008}
T.~Lappi, ``{Wilson line correlator in the MV model: Relating the glasma to
  deep inelastic scattering},''
  \href{http://dx.doi.org/10.1140/epjc/s10052-008-0588-4}{{\em Eur. Phys. J.}
  {\bfseries C55} (2008) 285--292},
\href{http://arxiv.org/abs/0711.3039}{{\ttfamily arXiv:0711.3039 [hep-ph]}}.

\bibitem{Fujii2009}
H.~Fujii, K.~Fukushima, and Y.~Hidaka, ``{Initial energy density and gluon
  distribution from the Glasma in heavy-ion collisions},''
  \href{http://dx.doi.org/10.1103/PhysRevC.79.024909}{{\em Phys. Rev.}
  {\bfseries C79} (2009) 024909},
\href{http://arxiv.org/abs/0811.0437}{{\ttfamily arXiv:0811.0437 [hep-ph]}}.

\bibitem{Fukushima2008}
K.~Fukushima, ``{Randomness in infinitesimal extent in the McLerran-Venugopalan
  model},'' \href{http://dx.doi.org/10.1103/PhysRevD.77.074005}{{\em Phys.
  Rev.} {\bfseries D77} (2008) 074005},
\href{http://arxiv.org/abs/0711.2364}{{\ttfamily arXiv:0711.2364 [hep-ph]}}.

\bibitem{GSL:2009}
B.~Gough, {\em GNU Scientific Library Reference Manual - Third Edition}.
\newblock Network Theory Ltd., 3rd~ed., 2009.

\bibitem{Fukushima:2006ax}
K.~Fukushima, F.~Gelis, and L.~McLerran, ``{Initial Singularity of the Little
  Bang},'' \href{http://dx.doi.org/10.1016/j.nuclphysa.2007.01.086}{{\em
  Nucl.Phys.} {\bfseries A786} (2007) 107--130},
\href{http://arxiv.org/abs/hep-ph/0610416}{{\ttfamily arXiv:hep-ph/0610416
  [hep-ph]}}.

\bibitem{Fukushima2012}
K.~Fukushima and F.~Gelis, ``{The evolving Glasma},''
  \href{http://dx.doi.org/10.1016/j.nuclphysa.2011.11.003}{{\em Nucl. Phys.}
  {\bfseries A874} (2012) 108--129},
\href{http://arxiv.org/abs/1106.1396}{{\ttfamily arXiv:1106.1396 [hep-ph]}}.

\bibitem{Lappi:2006fp}
T.~Lappi and L.~McLerran, ``{Some features of the glasma},''
  \href{http://dx.doi.org/10.1016/j.nuclphysa.2006.04.001}{{\em Nucl. Phys.}
  {\bfseries A772} (2006) 200--212},
\href{http://arxiv.org/abs/hep-ph/0602189}{{\ttfamily arXiv:hep-ph/0602189
  [hep-ph]}}.

\bibitem{Krasnitz2001}
A.~Krasnitz and R.~Venugopalan, ``{The Initial gluon multiplicity in heavy ion
  collisions},'' \href{http://dx.doi.org/10.1103/PhysRevLett.86.1717}{{\em
  Phys. Rev. Lett.} {\bfseries 86} (2001) 1717--1720},
\href{http://arxiv.org/abs/hep-ph/0007108}{{\ttfamily arXiv:hep-ph/0007108
  [hep-ph]}}.

\bibitem{Lappi2003}
T.~Lappi, ``{Production of gluons in the classical field model for heavy ion
  collisions},'' \href{http://dx.doi.org/10.1103/PhysRevC.67.054903}{{\em Phys.
  Rev.} {\bfseries C67} (2003) 054903},
\href{http://arxiv.org/abs/hep-ph/0303076}{{\ttfamily arXiv:hep-ph/0303076
  [hep-ph]}}.

\bibitem{Bodeker2007}
D.~Bodeker and K.~Rummukainen, ``{Non-abelian plasma instabilities for strong
  anisotropy},'' \href{http://dx.doi.org/10.1088/1126-6708/2007/07/022}{{\em
  JHEP} {\bfseries 07} (2007) 022},
\href{http://arxiv.org/abs/0705.0180}{{\ttfamily arXiv:0705.0180 [hep-ph]}}.

\bibitem{Edwards:2004sx}
{\bfseries SciDAC Collaboration, LHPC Collaboration, UKQCD Collaboration}
  Collaboration, R.~G. Edwards and B.~Joo, ``{The Chroma software system for
  lattice QCD},''
  \href{http://dx.doi.org/10.1016/j.nuclphysbps.2004.11.254}{{\em
  Nucl.Phys.Proc.Suppl.} {\bfseries 140} (2005) 832},
\href{http://arxiv.org/abs/hep-lat/0409003}{{\ttfamily arXiv:hep-lat/0409003
  [hep-lat]}}.

\bibitem{Iancu2003c}
E.~Iancu, K.~Itakura, and L.~McLerran, ``{A Gaussian effective theory for gluon
  saturation},'' \href{http://dx.doi.org/10.1016/S0375-9474(03)01477-5}{{\em
  Nucl. Phys.} {\bfseries A724} (2003) 181--222},
\href{http://arxiv.org/abs/hep-ph/0212123}{{\ttfamily arXiv:hep-ph/0212123
  [hep-ph]}}.

\bibitem{Fries2006}
R.~J. Fries, J.~I. Kapusta, and Y.~Li, ``{Near-fields and initial energy
  density in the color glass condensate model},''
\href{http://arxiv.org/abs/nucl-th/0604054}{{\ttfamily arXiv:nucl-th/0604054
  [nucl-th]}}.

\bibitem{Kovchegov2005}
Y.~V. Kovchegov, ``Can thermalization in heavy ion collisions be described by
  qcd diagrams?,''
  \href{http://dx.doi.org/10.1016/j.nuclphysa.2005.08.009}{{\em Nucl. Phys.}
  {\bfseries A762} (2005) 298--325}.

\bibitem{Lappi2006b}
T.~Lappi, ``{Energy density of the glasma},''
  \href{http://dx.doi.org/10.1016/j.physletb.2006.10.017}{{\em Phys. Lett.}
  {\bfseries B643} (2006) 11--16},
\href{http://arxiv.org/abs/hep-ph/0606207}{{\ttfamily arXiv:hep-ph/0606207
  [hep-ph]}}.

\bibitem{Baier:2000sb}
R.~Baier, A.~H. Mueller, D.~Schiff, and D.~T. Son, ``{'Bottom up'
  thermalization in heavy ion collisions},''
  \href{http://dx.doi.org/10.1016/S0370-2693(01)00191-5}{{\em Phys. Lett.}
  {\bfseries B502} (2001) 51--58},
\href{http://arxiv.org/abs/hep-ph/0009237}{{\ttfamily arXiv:hep-ph/0009237
  [hep-ph]}}.

\bibitem{Arnold:2002zm}
P.~B. Arnold, G.~D. Moore, and L.~G. Yaffe, ``{Effective kinetic theory for
  high temperature gauge theories},''
  \href{http://dx.doi.org/10.1088/1126-6708/2003/01/030}{{\em JHEP} {\bfseries
  01} (2003) 030},
\href{http://arxiv.org/abs/hep-ph/0209353}{{\ttfamily arXiv:hep-ph/0209353
  [hep-ph]}}.

\bibitem{Kurkela:2016vts}
A.~Kurkela, ``{Initial state of Heavy-Ion Collisions: Isotropization and
  thermalization},''
  \href{http://dx.doi.org/10.1016/j.nuclphysa.2016.01.069}{{\em Nucl. Phys.}
  {\bfseries A956} (2016) 136--143},
\href{http://arxiv.org/abs/1601.03283}{{\ttfamily arXiv:1601.03283 [hep-ph]}}.

\bibitem{Boguslavski:2018beu}
K.~Boguslavski, A.~Kurkela, T.~Lappi, and J.~Peuron, ``{Spectral function for
  overoccupied gluodynamics from real-time lattice simulations},''
  \href{http://dx.doi.org/10.1103/PhysRevD.98.014006}{{\em Phys. Rev.}
  {\bfseries D98} no.~1, (2018) 014006},
\href{http://arxiv.org/abs/1804.01966}{{\ttfamily arXiv:1804.01966 [hep-ph]}}.

\bibitem{Berges2014a}
J.~Berges, K.~Boguslavski, S.~Schlichting, and R.~Venugopalan, ``{Universal
  attractor in a highly occupied non-Abelian plasma},''
  \href{http://dx.doi.org/10.1103/PhysRevD.89.114007}{{\em Phys. Rev.}
  {\bfseries D89} no.~11, (2014) 114007},
\href{http://arxiv.org/abs/1311.3005}{{\ttfamily arXiv:1311.3005 [hep-ph]}}.

\end{thebibliography}\endgroup

\end{document}